\documentclass[aps,prb,preprint]{revtex4-2}  
\usepackage[utf8]{inputenc}

\usepackage{amsmath,amsthm,amssymb,amsfonts,dsfont}
\usepackage{siunitx,microtype,hyperref}
\usepackage[swedish,english]{babel}

\usepackage{graphicx,booktabs,rotating,colortbl,xcolor}
\usepackage{multirow}
\usepackage[caption=false]{subfig}

\usepackage{enumitem}

\DeclareMathOperator*{\argmin}{argmin}

\newcommand{\R}{\mathbb{R}}
\newcommand{\Prob}{\mathbb{P}}

\newcommand{\Id}{\mathds{1}}

\newcommand{\E}{\mathcal{E}}
\newcommand{\La}{\mathcal{L}}
\newcommand{\Rad}{\mathcal{R}}

\newcommand{\bA}{\boldsymbol{A}}
\newcommand{\ba}{\boldsymbol{a}}
\newcommand{\bc}{\boldsymbol{c}}
\newcommand{\bef}{\boldsymbol{f}}

\newcommand{\bw}{\boldsymbol{w}}

\newcommand{\x}{\boldsymbol{x}}
\newcommand{\y}{\boldsymbol{y}}

\newcommand{\bsigma}{\boldsymbol{\sigma}}
\newcommand{\bxi}{\boldsymbol{\xi}}

\newcommand{\bhata}{\boldsymbol{\hat a}}

\newcommand{\bhatf}{\boldsymbol{\hat f}}

\newcommand{\mueff}{\mu_{\text{eff}}}
\newcommand{\mueffi}{\mu_{\text{eff},\,i}}
\newcommand{\hatmueff}{\hat\mu_{\text{eff}}}

\begin{document}

	\title{Reconstructing Accurate Synthetic Hounsfield Units with Spectral CT Data}
	
	\author{Fredrik Grönberg}
	\email{gronberg@mi.physics.kth.se}
	\affiliation{Physics of Medical Imaging, KTH Royal Institute of Technology, AlbaNova University Center, SE-106 91 Stockholm, Sweden}

	\author{Hans Bornefalk}	
	\affiliation{\phantom{}Physics of Medical Imaging, KTH Royal Institute of Technology, AlbaNova University Center, SE-106 91 Stockholm, Sweden}

	\begin{abstract}
		\noindent
		\textbf{Purpose:} To evaluate a proposed method to reconstruct CT numbers that accurately mimic conventional CT numbers from spectral CT data, as would have been produced by a conventional system without effects of beam hardening.
		\\
		\\
		\textbf{Approach:} We implement the proposed method for simulated ideal and non-ideal photon counting multi-bin systems, the latter based on a photon counting Silicon detector, and compare them with a simulated ideal conventional energy integrating system with a cupping correction algorithm. We compare the systems using a mathematical phantom of a size and composition that produces severe cupping and beam hardening artefacts when imaged with a conventional system with no cupping correction.
		\\
		\\
		\textbf{Results:} The resulting images show CT numbers that are consistently accurate for a varying range of tissues and are free of beam hardening artefacts.
		\\
		\\ 
		\textbf{Conclusions:} This method could facilitate use of established rules-of-thumb regarding absolute CT numbers for various organs and conditions during the transition from conventional CT to spectral CT.	
		\end{abstract}

	\maketitle	  

\section{Introduction}

	Spectral computed tomography (CT) is a collective term for CT imaging techniques that acquire projection data at different effective energies. Various technical implementations exists that broadly fall in one of two categories: those where the multiple samples of the energy dependent line integrals of the linear attenuation coefficient are spatially perfectly aligned and those where they are not.

	Dual layer detector systems \cite{shefer2013state} and photon counting multi-bin systems \cite{shikhaliev2005photon, schlomka2008experimental, iwanczyk2009photon, bornefalk2010photon, kappler2012first} belong to the first category and therefore allow material basis decomposition in the projection domain, a vital step for acquiring quantitative CT images, iodine maps and synthetic mono-energetic images \cite{alvarez1976energy,alvarez1979comparison,lehmann1981generalized,roessl2007k}. Dual source methods \cite{flohr2006first} and switched kVp \cite{kalender1986evaluation} belong to the latter category. Both use standard scintillator detector technology and apply (two) different acceleration voltages for different projection lines. This way, two somewhat under-sampled and misaligned sinograms at two different effective energies are generated. In rapid kVp-switching systems, the misalignment is only one detector pitch and interpolation of the data, associated with a small cost in spatial resolution, makes it possible to obtain two effective projection estimates at different energies for each projection line and generate beam hardening free synthetic mono-energetic images \cite{goodsitt2011accuracies}. In dual source systems, where two x-ray tubes and detectors offset by 90$^\circ$ are used, the misalignment is larger and any analysis attempting to make use of the difference in linear attenuation coefficients between different tissues must be carried out in the image domain after reconstruction \cite{taguchi2007image}, making accurate quantitative CT more difficult.

	Regardless of which multi-energy CT technology is used, i.e. if the spectral information is obtained from the projection domain or the image domain, it is desirable to reduce the data and generate a single cross sectional image. This single image should be easily interpretable to the radiologist, and allow her to make an initial assessment and to direct her attention to certain areas that can be more thoroughly scrutinized, e.g. in a synthetic mono-energetic image. For multi-energy systems operating in the image domain, this is typically achieved by blending the high- and low-energy images \cite{holmes2008evaluation,kim2010image}. For systems where material basis decomposition can be performed in the projection domain, synthetic mono-energetic images displayed at an average transmitted x-ray energy are typically used \cite{goodsitt2011accuracies}. Both of these systems are however unable to accurately reconstruct the Hounsfield units (also known as CT numbers) generated by an energy integrating system simultaneously for all relevant types of tissue.

	We have earlier proposed a method to generate synthetic Hounsfield units that mimic those of energy integrating systems \cite{bornefalk2012synthetic}. Unlike synthetic mono-energetic images, the method produces CT numbers that are accurate for all tissues. In this paper we implement that method on mathematical phantoms of sufficient size and composition to produce severe cupping and beam hardening artifacts when imaged with a simulated conventional system; despite this the method is capable to generate not only beam hardening free images, but also CT numbers that for all materials appear as if imaged with a conventional CT scanner. We implement the method for both an ideal and a non-ideal photon counting multi-bin system, where the non-ideal system is based on a photon counting Silicon detector \cite{bornefalk2010photon}. To provide a fair comparison with the simulated conventional system, we have implemented the empirical cupping correction (ECC) of \cite{kachelriess2006empirical} in its image reconstruction.

\section{Background}

	In conventional CT, a linear attenuation coefficient (LAC) distribution $\mu(\boldsymbol{x},\E)$ is imaged by performing tomographic reconstruction on a set of projection values
	\begin{equation}\label{eq:projection_data}
		p[\mu] \triangleq \left\{p_i[\mu]\right\}_{i=1}^N, \quad p_i[\mu] \triangleq -\log\frac{I(\ell_i)}{I_0(\ell_i)},
	\end{equation}
	where $\ell_i$ denotes the $i$th projection line, $I(\ell_i)$ the attenuated signal and $I_0(\ell_i)$ the unattenuated signal in that line. We take the projection values to be line integrals of an \emph{effective} LAC distribution $\mueffi(\x)$ specific to the projection line,
	\begin{equation}
		p_i[\mu] = \int_{\ell_i}\mueffi(\x)\, ds.
	\end{equation}
	As was shown in \cite{bornefalk2012synthetic}, in the absence of noise and object scatter it follows from the mean value theorem of integrals that
	\begin{equation}
		\mueffi(\x) = \mu(\x,\E_i'),
	\end{equation}
	where the effective energy $\E_i'$ depends on the projection line due to the physical process of beam hardening. The set $p[\mu]$ is a thus a discrete Radon transform of the LAC distribution $\mu$ evaluated at the energies $\E_i'$ for each projection line. We let $\hatmueff(\x)$ denote the reconstructed image
	\begin{equation}
		\hatmueff(\x) \triangleq \Rad^{-1}p[\mu](\x),
	\end{equation}
	where $\Rad^{-1}$ is some implementation of the inverse discrete Radon transform. Beam hardening and cupping artefacts result from the fact that $\E_i'$ is not constant for all projection lines; specifically $\E_i'$ is higher for projection lines which are more strongly attenuated. The Hounsfield unit image is defined as
	\begin{equation}\label{eq:HU}
		\text{HU}(\x) \triangleq 1000\,\frac{\hatmueff(\x) - \hatmueff^{\text{water}}}{\hatmueff^{\text{water}} - \hatmueff^{\text{air}}},
	\end{equation}
	where $\hatmueff^{\text{air}}$ and $\hatmueff^{\text{water}}$ are the system specific effective LACs of air and water, presumably in the beam hardening free limit.

	\subsection{Forward Models}

		We limit our study to quantum noise limited systems, thus disregarding effects of object scatter and electronic noise in the forward models. A statistical model for a conventional CT system in this case is that $I(\ell_i)$ in \eqref{eq:projection_data} is a Gaussian random variable with mean $\nu_i$ and and variance $\sigma_i^2$ given by
		\begin{equation}\label{eq:conventional_forward_model}
			\begin{split}
					\nu_{i} &= \int_\R N_{i}(\E)\,\E\exp\left(-\int_{\ell_i}\mu(\x,\E)\,ds\right)d\E, \\
					\sigma^2_{i} &= \int_\R N_{i}(\E)\,\E^2\exp\left(-\int_{\ell_i}\mu(\x,\E)\,ds\right)d\E.
			\end{split}
		\end{equation}
		A photon counting system is modeled by assuming that the number of counts in the $k$th energy bin in the $i$th projection line $\ell_i$ is a Poisson random variable with mean
		\begin{equation}\label{eq:forward_model}
			\lambda_{ik} = \int_\R N_{i}(\E)S_k(\E)\exp\left(-\int_{\ell_i}\mu(\x,\E)\,ds\right)d\E.
		\end{equation}
		For both types of system, $N_{i}(\E)$ denotes the product of the incident spectrum and the detection efficiency of the detector,
		\begin{equation}\label{eq:w}
			N_{i}(\E) = I_{0,i}\,\Phi_{i}(\E)D(\E),
		\end{equation}
		where $I_{0,i}$ denotes the number of incident photons in the projection line, $\Phi_{i}(\E)$ its normalized energy distribution and $D(\E)$ the detection efficiency of the detector material. For the spectral system, $S_k(\E)$ denotes the $k$th bin response function, i.e. the probability that a detected photon with energy $\E$ is counted in the $k$th energy bin, a simple model of which is
		\begin{equation}\label{eq:bin_response}
			S_k(\E) = \int_{T_{k-1}}^{T_k}R(\E,\E')\,d\E', \quad k = 1,\dots,K,
		\end{equation}
		where $T_{k-1}$ and $T_{k}$ are the bin edges of the $k$th energy bin and $R(\E,\E')$ is the detector response function, i.e. the probability that detected photon with energy $\E$ will deposit an energy $\E'$ in the detector. A more detailed analysis of these forward models is found in \cite{roessl2009cramer}. In the following we let the product of $N_i(\E)$ and a particular system response function be denoted by $w(\E)$, i.e.
		\begin{equation}\label{eq:w_i}
			\begin{split}
			w_i(\E) &\triangleq N_i(\E)\,\E, \\
			w_{ik}(\E) &\triangleq N_i(\E) S_k(\E).
			\end{split}
		\end{equation}

	\subsection{Basis Decomposition}

		Under the assumption of $Z$-$\E$ separability of photon interaction cross sections, the LAC distribution $\mu(\boldsymbol{x},\E)$ may be decomposed into $L$ spatially dependent basis coefficients $a_l(\x)$ and energy dependent basis functions $f_l(\E)$ \cite{alvarez1976energy},
		\begin{equation}\label{eq:decomposition}
			\mu(\boldsymbol{x},\E) = \sum_{l=1}^L a_l(\x)f_l(\E).
		\end{equation}
		For compactness of notation, we define
		\begin{equation}
			\begin{split}
			\ba(\x) &\triangleq (a_1(\x)),\dots,a_L(\x))^T \in \R^L, \\
			\bef(\E) &\triangleq (f_1(\E),\dots,f_L(\E)) \in \R^{1\times L},
			\end{split}
		\end{equation}
		where the superscript $T$ denotes the vector transpose. With this convention of dimensions, a discretization of $\bef(\E)$ has a discretized basis function in each column. We define the line integral of $a_l(\x)$ along $\ell_i$ and the vector of such line integrals as
		\begin{equation}
			A_{il} \triangleq \int_{\ell_i}a_l(\x)\,ds, \quad \bA_i \triangleq (A_{i1},\dots,A_{iL})^T \in \R^L.
		\end{equation}
		Inserting (\ref{eq:decomposition}) and \eqref{eq:w_i} into (\ref{eq:forward_model}) then yields the following parametrization of $\lambda_{ik}$,
		\begin{equation}\label{eq:parametrized_model}
			\lambda_{ik}(\bA_i) = \int_\R w_{ik}(\E)e^{-\bef(\E)\bA_i}d\E.
		\end{equation}
		Let $\y_i = (y_{i1},\dots,y_{iK})$ denote observed signals in $\ell_i$. Under the assumption that the signals are outcomes of independent Poisson random variables, the parametrization \eqref{eq:parametrized_model} can be used to find a maximum likelihood estimate of $\bA_i$ \cite{roessl2007k}. The probability of having observed $\y_i$, given the set of line integrals $\bA_i$ is given by
		\begin{equation}
			\Prob(\y_i\,|\bA_i) = \prod_{k=1}^K \frac{\lambda_{ik}(\bA_i)^{y_{ik}}}{y_{ik}!}\exp\left(-\lambda_{ik}(\bA_i)\right)
		\end{equation}
		and the corresponding negative log-likelihood of an arbitrary $\bA$, given the set of observed signals $\y_i$ is obtained by taking the negative logarithm of this probability and discarding terms that are constant with respect to $\bA$,
		\begin{equation}
			\La_i(\bA) \triangleq \sum_{k=1}^K \lambda_{ik}(\bA) - y_{ik}\log\lambda_{ik}(\bA).
		\end{equation}
		The minimizer of $\La_i(\bA)$ is the set of line integrals for which the probability of observing $\y_i$ is maximal, and we define the maximum likelihood estimator of $\bA_i$ as
		\begin{equation}\label{eq:ML}
			\bA_i^* \triangleq \argmin_{\substack{\bA\in\R^L \\ \bA^T\bef(\E)\,\geq\, 0}} \La_i(\bA),
		\end{equation}
		where the non-negativity constraint implies the non-negativity of the LAC. Basis decomposition is performed by solving (\ref{eq:ML}) for $i = 1,\dots,N$. Let $A_l^* \triangleq \{A_{il}^*\}_{i=1}^N$ denote the resulting set of line integral estimates, where $A_{il}^*$ is the $l$th component of $\bA_i^*$.
		Basis image estimates are then given by
		\begin{equation}\label{eq:image_estimates}
			\begin{split}
				\bhata(\x) &\triangleq (\hat a_1(\x),\dots,\hat a_L(\x))^T \in \R^L, \\
				\hat a_l(\x) &\triangleq \Rad^{-1}A_l^*(\x), \quad l = 1,\dots,L.
			\end{split}
		\end{equation}
		A beam hardening free estimate of the LAC distribution $\mu(\x,\E)$ is finally given by
		\begin{equation}\label{eq:reconstructed_LAC}
			\hat\mu(\x,\E) = \sum_{l=1}^L \hat a_l(\x)f_l(\E) = \bef(\E)\,\bhata(\x)
		\end{equation}
		and a so-called synthetic mono-energetic image is obtained by evaluating \eqref{eq:reconstructed_LAC} at a certain energy $\E'$. A mono-energetic Hounsfield unit image is obtained by applying \eqref{eq:HU} to such an image.

	\subsection{Ideal Hounsfield Units}\label{sec:ideal_units}

		For the purpose of self-containment we present a derivation of \emph{ideal} Hounsfield units in the sense of being beam hardening free, as originally done in \cite{bornefalk2012synthetic}, with slightly more rigour. Assume a one-dimensional system for which $p[\mu]$ as given by \eqref{eq:projection_data} is a single projection value for a projection line $\ell$ and further that we are in the limit where $I(\ell)$ and $I_0(\ell)$ are equal to their means as given by \eqref{eq:conventional_forward_model}. We make the following parametrization of the line $\ell$ with length $\Delta\ell$,
		\begin{equation}
			\ell = \left\{
				 \begin{gathered}
				 	\x + t\,\bxi \: : \: \|\bxi\| = 1, \\
				 	a\Delta\ell\leq t\leq b\Delta\ell, \; a < 0 <  b, \; b - a = 1
				 \end{gathered}\right\},
		\end{equation}
		for which $\x\in\ell$ for any $\Delta\ell \geq 0$ and
		\begin{equation}
			\int_{\ell} f \,ds = \int_{a\Delta\ell}^{b\Delta\ell} f(\bsigma(t))\,dt, \quad \bsigma(t) = \x + t\,\bxi.
		\end{equation}
		As the sample thickness tends to zero, the effective LAC $\mueff(\x)$ becomes beam hardening free, and so we define the ideal effective LAC as
		\begin{equation}\label{eq:ideal_mu}
			\mueff^*(\x)\triangleq\lim_{\Delta\ell\rightarrow 0} \frac{p[\mu]}{\Delta\ell}.
		\end{equation}
		To compute \eqref{eq:ideal_mu} we will assume that $\mu$ is Lipschitz continuous in $\x$ for $\x$ in $\ell$, a plausible assumption if the sample is taken to be a single type of tissue. Let $C$ denote the implied Lipschitz constant. It then holds that
		\begin{equation}
			\int_{a\Delta\ell}^{b\Delta\ell}\mu(\x + t\,\bxi,\E)\,dt = \Delta\ell\,\mu(\x,\E) + o(\Delta\ell),
		\end{equation}
		where $o(\Delta\ell)$ comprises terms that tend to zero faster than $\Delta\ell$ as $\Delta\ell$ tends to zero.
		\begin{proof}
			By the mean value theorem of integrals, there is a $c\in[a\Delta\ell,b\Delta\ell]$ and a corresponding $\bc = \x + c\bxi\in\ell$ such that
			\begin{equation}
				\int_{a\Delta\ell}^{b\Delta\ell}\mu(\x + t\,\bxi,\E)\,dt = \Delta\ell \,\mu(\bc,\E).
			\end{equation}
			By the Lipschitz continuity of $\mu$ in $\x$ it then holds that
			\begin{equation}
				\begin{split}
					&\quad\left|\, \Delta\ell\,\mu(\x,\E) - \Delta\ell \,\mu(\bc,\E)\,\right| = \Delta\ell\left|\,\mu(\x,\E) - \mu(\bc,\E)\,\right| \\
					&\leq C\Delta\ell \, \|\x - \bc \| \leq C(\Delta\ell)^2 = o(\Delta\ell)
				\end{split}
			\end{equation}
			and the desired conclusion follows.
		\end{proof}
		Using this result, we have by \eqref{eq:projection_data}, \eqref{eq:conventional_forward_model} and \eqref{eq:w_i} that
		\begin{equation}
			\begin{split}
			p[\mu] 	&= -\log\frac{I(\ell)}{I_0(\ell)} \\
					&= -\log\frac{\int_\R w(\E)\exp\left(-\int_{a\Delta\ell}^{b\Delta\ell}\mu(\x + t\,
						\bxi,\E)\,dt\right)d\E}{\int_\R w(\E)\,d\E} \\
				   	&= -\log \frac{\int_\R w(\E)\exp\left(-\Delta\ell\,\mu(\x,\E) + o(\Delta\ell)\right)	d\E}{\int_\R w(\E)\,d\E} \\
					&= -\log \frac{\int_\R w(\E)\left(1 - \Delta \ell\,\mu(\x,\E) + o(\Delta\ell)\right)	d\E}{\int_\R w(\E)\,d\E} \\
					&= -\log \left(1 - \Delta\ell \,\frac{\int_\R w(\E)\mu(\x,\E)\,d\E}{\int_\R w(\E)\,		d\E} + o(\Delta\ell)\right) \\
					&= \Delta\ell\, \frac{\int_\R w(\E)\mu(\x,\E)\,d\E}{\int_\R w(\E)\,d\E} + o(\Delta\ell),
			\end{split}
		\end{equation}
		where we use the standard expansions
		\begin{equation}
				e^x = 1 + x + o(x), \quad \log(1 + x) = x + o(x).
		\end{equation}
		Thus
		\begin{equation}
			\frac{p[\mu]}{\Delta\ell} = \frac{\int_\R w(\E)\mu(\x,\E)\,d\E}{\int_\R w(\E)\,d\E} + o(1)
		\end{equation}
		and in the limit $\Delta\ell\rightarrow0$ it follows that
		\begin{equation}\label{eq:ideal_LAC}
			\mu_{\text{eff}}^*(\x) = \frac{\int_\R w(\E)\mu(\x,\E)\,d\E}{\int_\R w(\E)\,d\E} \triangleq\langle \mu(\x,\cdot)\rangle_w,
		\end{equation}
		where $\langle \cdot\rangle_w$ denotes the weighted mean in the energy variable with the weighting function $w(\E)$. The ideal Hounsfield unit of a certain tissue follows from inserting its ideal effective LAC in \eqref{eq:HU}.

		The main concept of \cite{bornefalk2012synthetic} and this paper is that that the ideal effective LAC can be reconstructed using a specific weighting of the basis images $a_l(\x)$. Specifically, let
		\begin{equation}\label{eq:f_w}
			\bhatf_w \triangleq (\langle f_1\rangle_w,\dots,\langle f_L\rangle_w) \in \R^{1\times L},
		\end{equation}
		it then follows by \eqref{eq:decomposition} and \eqref{eq:ideal_LAC} that
		\begin{equation}\label{eq:effective_decomposition}
			\begin{split}
			\mueff^*(\x) &= \Big\langle\sum_{l=1}^L a_l(\x)f_l(\cdot)\Big\rangle_w \\ &= \sum_{l=1}^L a_l(\x)\langle f_l\rangle_w = \bhatf_w\ba(\x)
			\end{split}
		\end{equation}
		which suggests that an ideal Hounsfield unit image can be constructed by weighting the estimated basis images $\bhata(\x)$, as given by \eqref{eq:image_estimates}, the same way. Analogous to \eqref{eq:HU}, we define the synthetic Hounsfield unit image as
		\begin{equation}\label{eq:SHU}
			\text{SHU}(\x) \triangleq 1000\frac{\bhatf_w\bhata(\x) - \bhatf_w\bhata^{\text{water}}}{\bhatf_w\bhata^{\text{water}} - \bhatf_w\bhata^{\text{air}}},
		\end{equation}
		where $\bhata^{\text{water}}$ and $\bhata^{\text{air}}$ are estimated basis coefficient vectors of water and air.

		\subsection{Polychromaticity of the Ideal Hounsfield Unit}

		By another application of the mean value theorem of integrals, it follows that the ideal effective LAC is equal to the LAC evaluated at an \emph{ideal effective energy} $\E^*$,
    \begin{equation}\label{eq:effective_energy}
      \mu(\E^*) = \langle \mu\rangle_w
    \end{equation}
    and similarly, that the basis image weighting that reproduces the ideal effected LAC, $\bhatf_w$, is equal to $\bef(\E)$ evaluated at a set of energies $\E_1^*,\dots,\E_L^*$ such that
		\begin{equation}
			f_l(\E^*_l) = \langle f_l\rangle_w, \quad l = 1,\dots,L.
		\end{equation}
    The difference is that the energy $\E^*$ is material dependent whereas the energies $\E_1^*,\dots,\E_L^*$ are not. This explains the failure of Hounsfield units created from synthetic monoenergetic data to reproduce accurate values consistently for multiple types of tissue at any single reconstruction energy \cite{goodsitt2011accuracies}, hence the need to consider polychromatic formations of basis images \cite{bornefalk2012synthetic}.

\section{Methods}

	\subsection{Phantom}\label{sec:phantom}
\begin{figure}[htbp]
			\centering
			\begin{minipage}[b]{\linewidth}
				\subfloat[]{\includegraphics[width=0.5\linewidth]{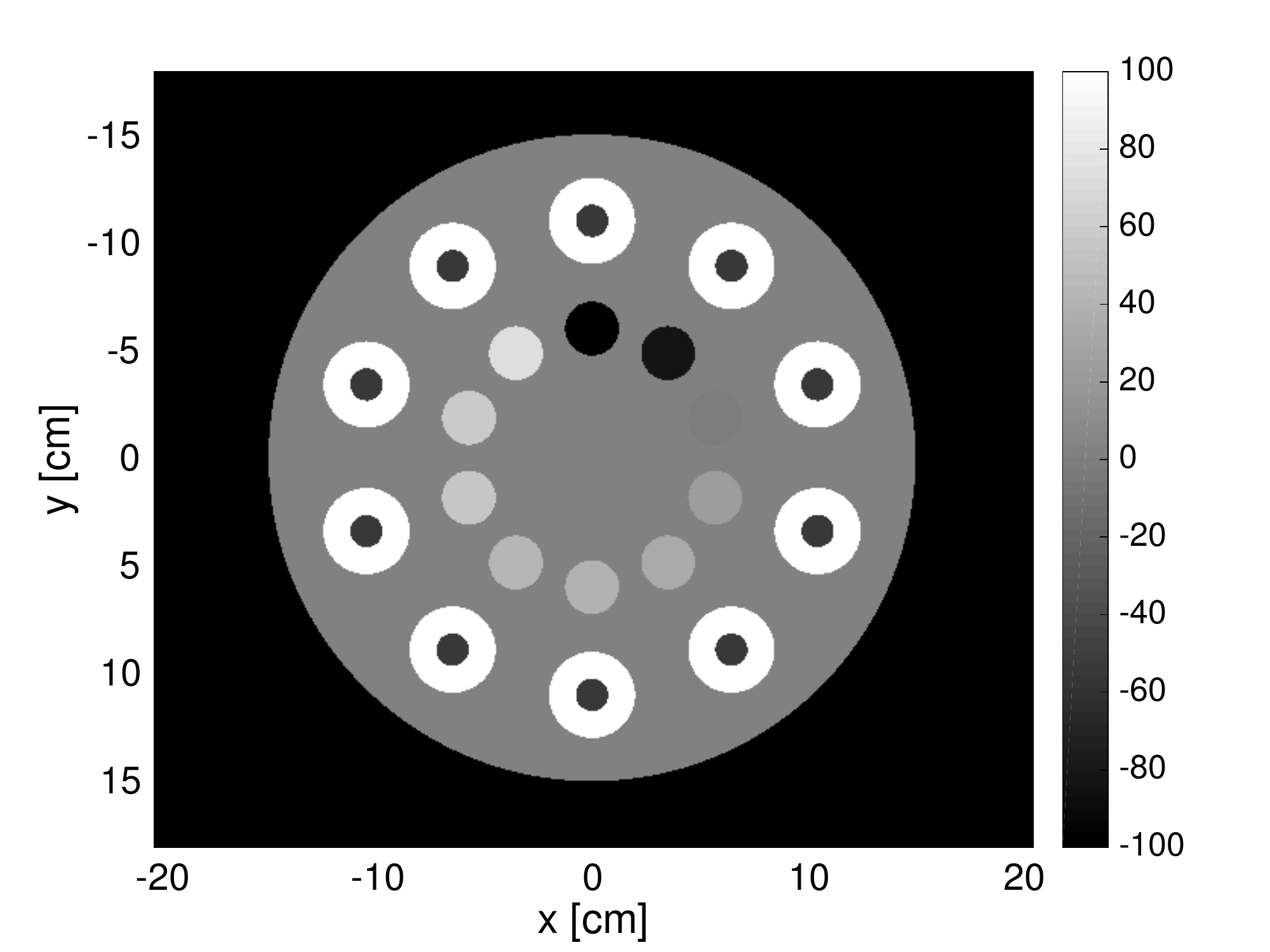}}				
				\hfil
				\subfloat[]{\includegraphics[width=0.5\linewidth]{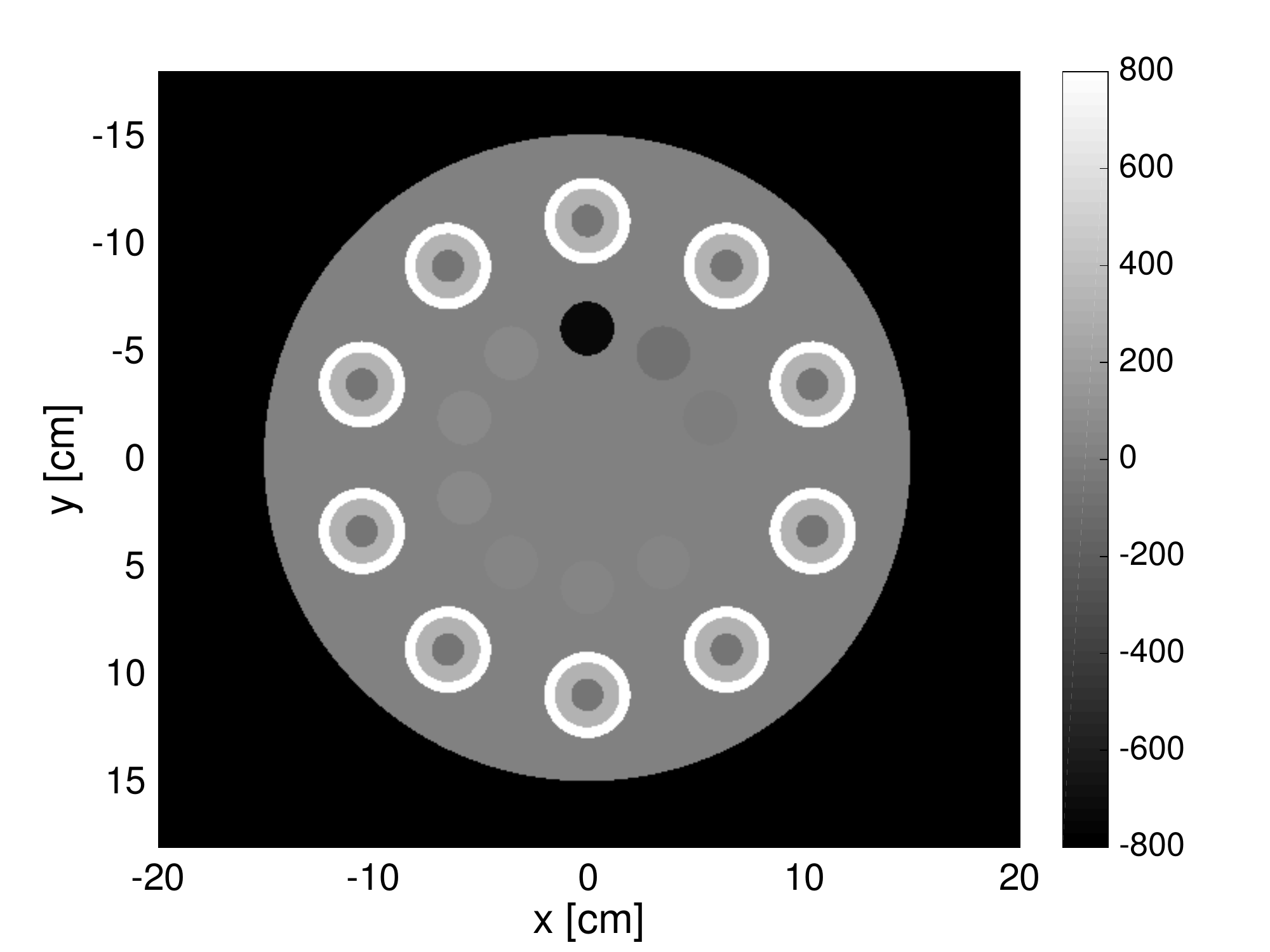}}			
				\vfil
				\subfloat[\label{fig:roi}]{\includegraphics[width=0.5\linewidth]{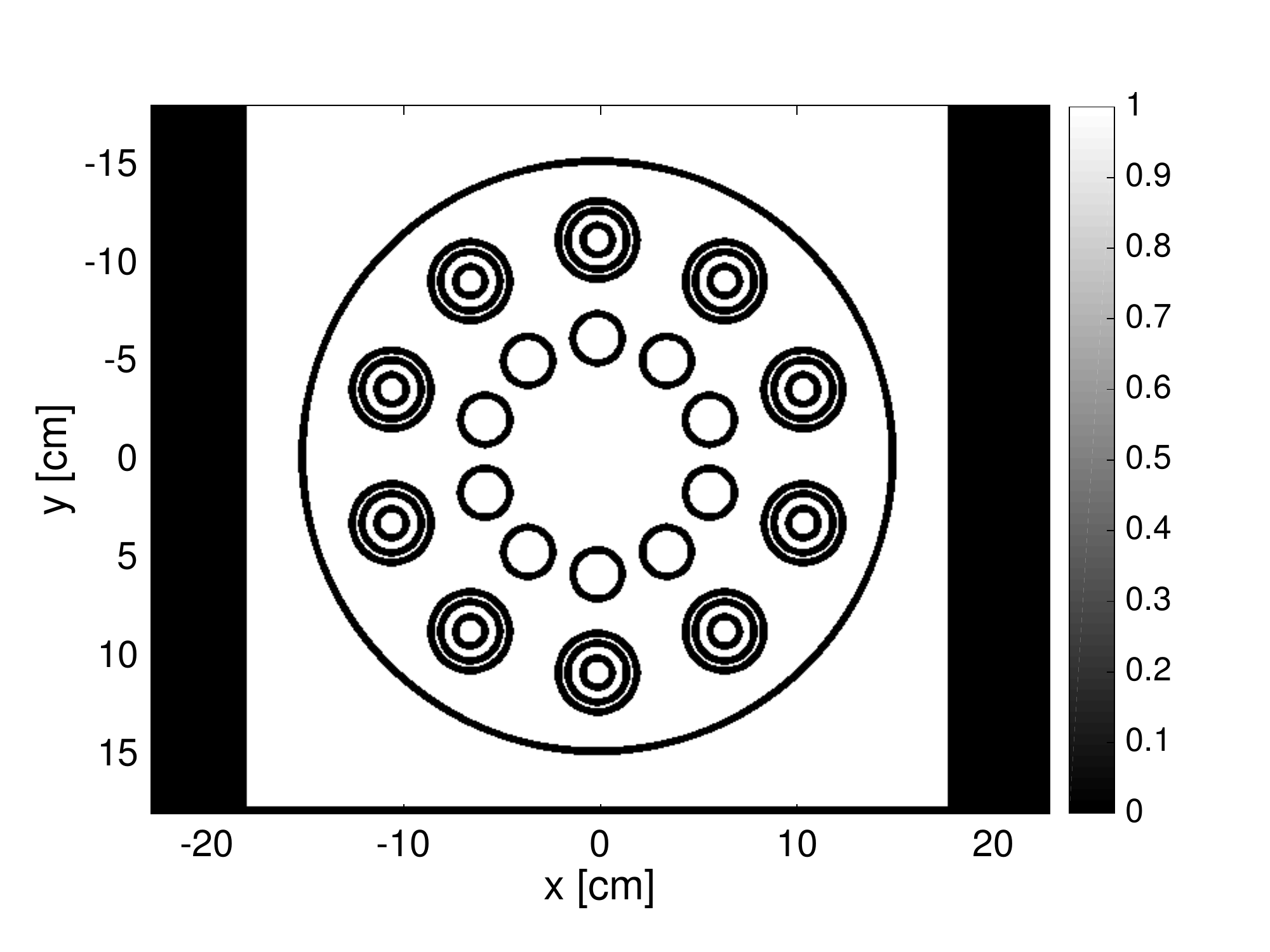}}
			\end{minipage}
			
			\caption{(a) Validation phantom displayed in ideal Hounsfield units with a window of $[-100,100]$. (b) Validation phantom displayed in ideal Hounsfield units with a window of $[-800,800]$. (c) Regions of interest in the validation phantom (in white). Boundary regions (in black) are not included when computing the figure of merit. The materials in the outer array of inserts are, in order from the outer to the inner layer of each insert, as presented in the second section of Tab. \ref{tab:materials}. The materials in the inner array of inserts are, in clockwise order starting from the top, as presented in the third section of Tab. \ref{tab:materials}.}\label{fig:phantom}
		\end{figure}

		We create two simulated phantoms: a water phantom for calibration and a validation phantom with two concentrical, circular arrays of circular inserts. Both phantoms are circular with 15 cm radius. In the validation phantom, the inserts in the outer array consist of a 5 mm layer of bone, a 7.5 mm layer of spongiosa and 7.5 mm layer of yellow marrow, with a total radius of 2 cm. The inserts in the inner array have a radius of 1.25 cm and contain a range of soft tissues. The surrounding material in the phantom is water and the region outside the phantom is air. The LACs of the materials are created using elemental LAC data from the XCOM database \cite{berger2013xcom} and mixture coefficients from the ICRU-44 \cite{white1989tissue}. To compute an ideal Hounsfield unit image of the phantom, we use \eqref{eq:HU} and \eqref{eq:ideal_LAC} with the weighting function
		\begin{equation}\label{eq:ideal_weight}
			w(\E) = \Phi(\E)\,\E,
		\end{equation}
		where the source x-ray spectrum $\Phi(\E)$ is generated using the x-ray tube model of \cite{cranley1997catalogue} assuming 120 kVp tungsten spectrum, $7^\circ$ anode angle and filtration by 3 mm aluminum and 0.5 mm copper. Images of the validation phantom, displayed in ideal Hounsfield units, are presented in Fig. \ref{fig:phantom}. The phantom materials, their ideal Hounsfield units and the corresponding ideal effective energies as given by \eqref{eq:effective_energy} are presented in Tab. \ref{tab:materials}.

	   \begin{table}[htbp]
			\centering
			\begin{tabular}{r | r | c}
			Material 		& Ideal HU & $\E^*$ [keV]	\\
			\hline
			Air  					& -1000 &	69 \\
			Water 				& 0   	&	70 \\
			\hline
			Yellow marrow & -54		& 71 \\
			Spongiosa 		& 321   & 68 \\
			Bone 					& 1701 	& 67 \\
			\hline
			Inflated Lung & -742	&	69 \\
			Adipose 			& -81   &	71 \\
			Breast 				& -2 		& 70 \\
			GI Tract 			& 23 		& 70 \\
			Pancreas 			& 32 		& 70 \\
			Brain 				& 49 		& 69 \\
			Muscle 				& 44 		& 69 \\
			Liver 				& 54 		& 69 \\
			Blood 				& 57 		& 69 \\
			Skin 					& 73    & 70
			\end{tabular}\caption{Materials in the validation phantom, their corresponding ideal Hounsfield units and ideal effective energies.}\label{tab:materials}
		\end{table}

	\subsection{Projection Data}

		Projection data for both phantoms is created for three detector systems: an ideal conventional energy integrating system modeled with \eqref{eq:conventional_forward_model} and two photon counting multi-bin systems modeled with \eqref{eq:forward_model}, one of them ideal and the other modeled after a photon counting Silicon detector studied in \cite{bornefalk2010photon}. All systems are assumed to have a fan-beam geometry with 50 cm source-to-isocenter distance, 100 cm source-to-detector distance and 1 mm $\times$ 1 mm detector pixels.

		To model $N_i(\E)$ as given by \eqref{eq:w}, we start from the central projection line. In this line, the x-ray spectrum $\Phi_i(\E)$ is generated assuming the same parameters used in \eqref{eq:ideal_weight} to compute the ideal Hounsfield units in Tab. \ref{tab:materials}. In order to simulate the effects of a bowtie-filter, additional aluminum filtration is added to the non-central projection lines in such a way that all projection lines have the same number of photons after transmission through the water phantom. The number of photons in each projection line, $I_{0,i}$, is then computed with a current-time product such that the absorbed dose in the water phantom equals 25 mGy and is then multiplied by a factor corresponding to an assumed number of added slices: ten slices of the validation phantom and forty slices of the calibration phantom. With the assumed magnification factor this corresponds to a 5 mm and a 20 mm slice in isocenter, respectively. The system specific detection efficiency, bin response functions and projection data are then created as follows.

		\subsubsection{Ideal Conventional Energy Integrating System}

			The detection efficiency is assumed to be $D(\E) = 1$. Gaussian distributed projection data is then created using \eqref{eq:conventional_forward_model}.

		\subsubsection{Ideal Photon Counting Multi-Bin System}

			The detection efficiency is assumed to be $D(\E) = 1$ and the detector response function $R(\E,\E') = \delta(\E - \E')$. The resulting bin response functions $S_k(\E)$, given by \eqref{eq:bin_response}, are then $\Id_{[T_{k-1},T_k)}(\E)$ where $\Id_\chi$ denotes the indicator function of the set $\chi$. The system is assumed to have five energy bins with thresholds set to produce an approximately equal number of photons in each bin for the calibration phantom. Poisson distributed projection data is then created using \eqref{eq:forward_model}.

		\subsubsection{Non-Ideal Photon Counting Multi-Bin System}

			This system is modeled after a photon counting Silicon detector, previously studied in \cite{bornefalk2010photon}. The detector is assumed to have a dead layer of depth $d_{\text{d}} = 0.05$ mm and an active layer of depth $d_{\text{a}} = 30$ mm, which yields a detection efficiency
			\begin{equation}
				D(\E) = e^{-\mu_{\text{Si}}(\E)d_{\text{d}}}(1 - e^{-\mu_{\text{Si}}(\E)d_{\text{a}}}),
			\end{equation}
			where $\mu_{\text{Si}}(\E)$ is the LAC of Silicon. The detecor response function is modeled as
			\begin{equation}
				R(\E,\E') = \left(\omega_{\text{Ph}}(\E)\delta(\E - \E') + \omega_{\text{Co}}(\E)R_{\text{Co}}(\E,\E')\right)*f_\sigma(\E),
			\end{equation}
			where $\omega_{\text{Ph}}(\E)$ and $\omega_{\text{Co}}(\E)$ are the relative interaction cross sections of photoelectric effect and Compton scattering for Silicon, obtained from the XCOM database \cite{berger2013xcom}, $R_{\text{Co}}(\E,\E')$ the probability that a photon of energy $\E$ deposits an energy $\E'$ in a Compton event, derived from the Klein-Nishina formula \cite{klein1929streuung},	and $f_\sigma(\E)$ the Gaussian density function with standard deviation $\sigma = 1.5$ keV, corresponding to the measured energy resolution of a prototype Silicon detector operating at a moderate count rate \cite{xu2013evaluation}. This system is also assumed to have five energy bins, with thresholds set to produce an approximately equal number of photons in each bin for the calibration phantom. The resulting bin response functions are computed according to \eqref{eq:bin_response} and Poisson distributed projection data is then created using \eqref{eq:forward_model}.

\begin{figure}[htbp]
				\centering
				\begin{minipage}[b]{\linewidth}
					\subfloat[]{\includegraphics[width=0.5\linewidth]{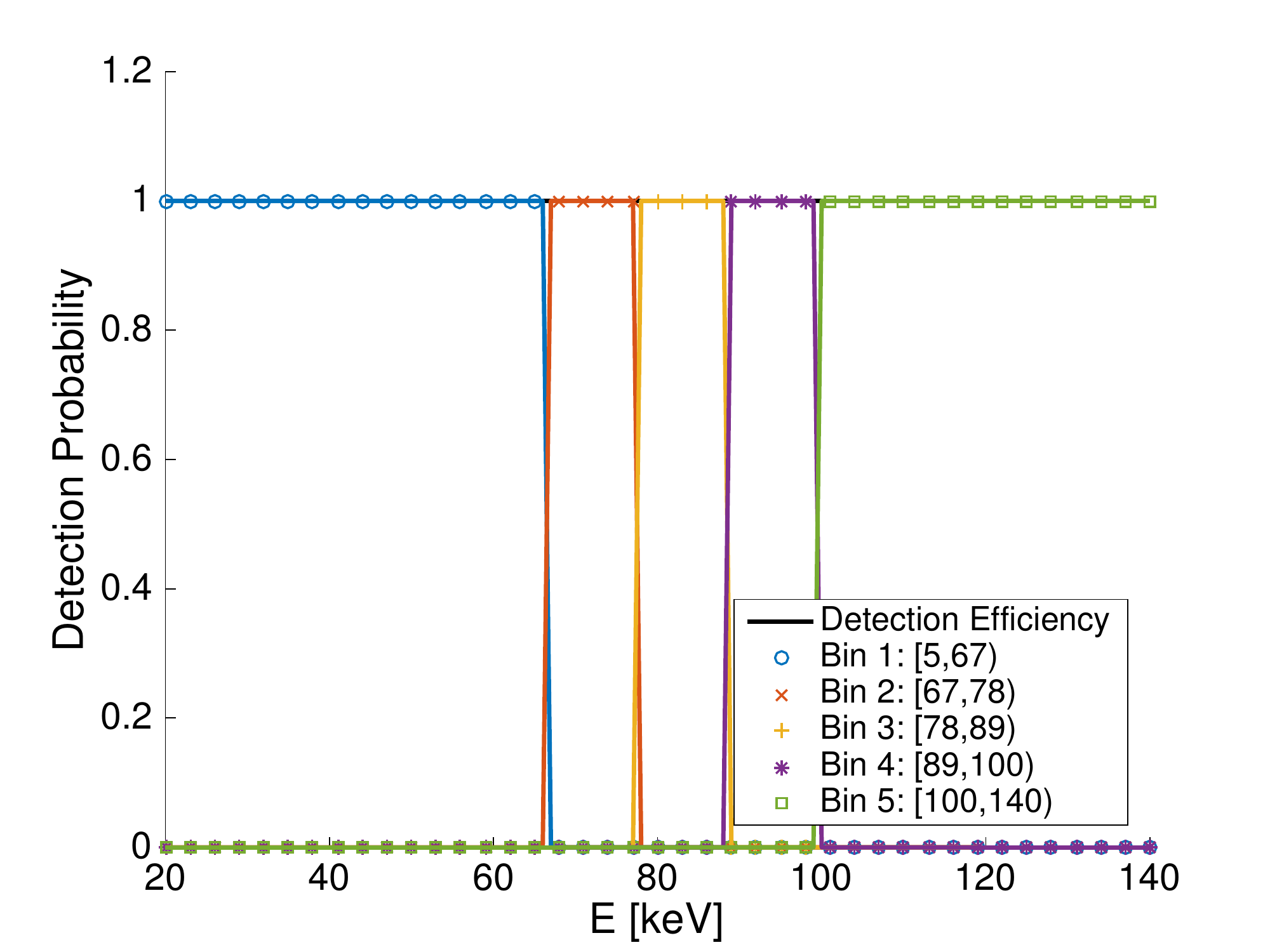}}
					\subfloat[]{\includegraphics[width=0.5\linewidth]{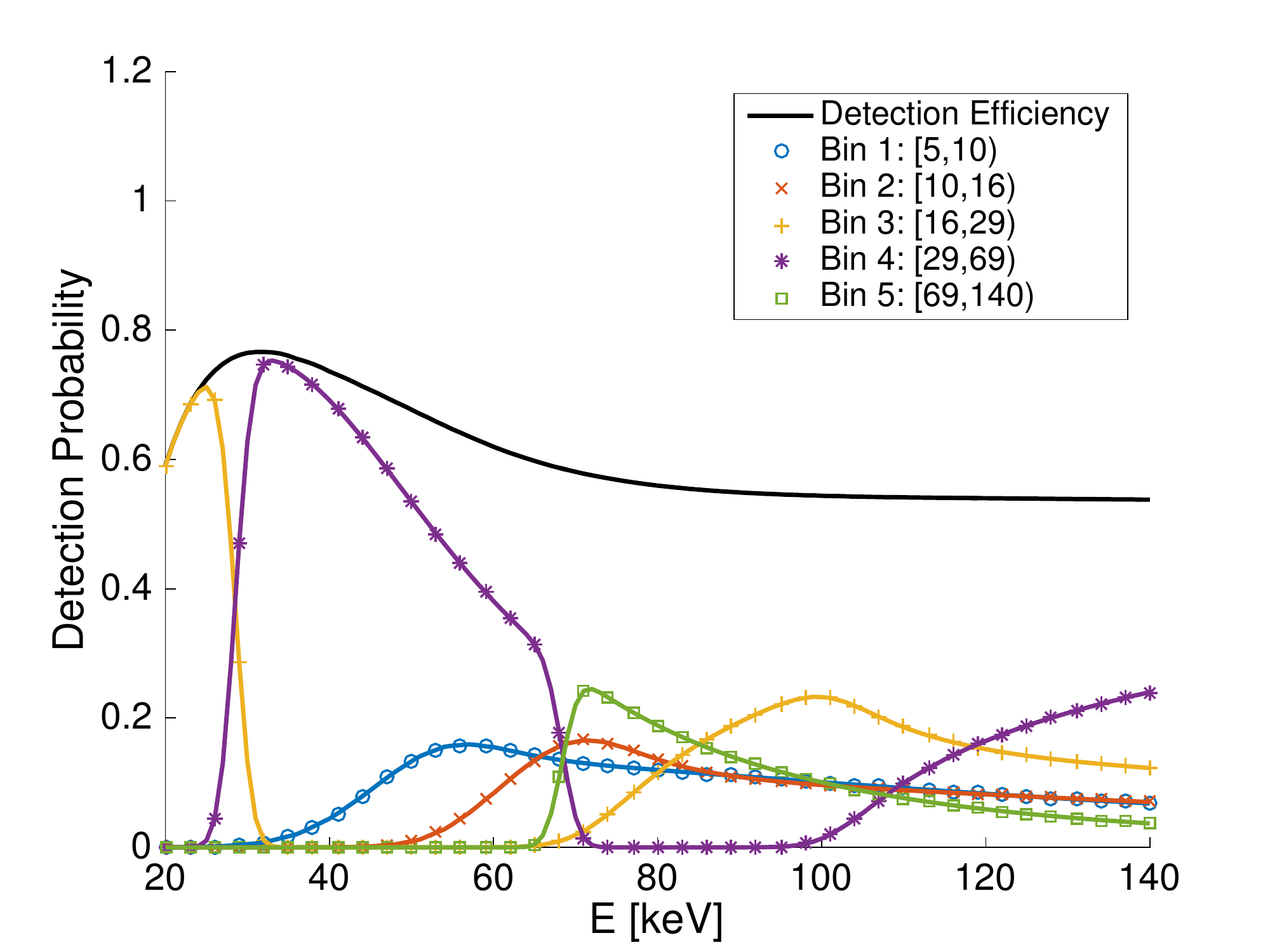}}					
				\end{minipage}								
				\caption{Product of detection efficiency and bin response functions for the (a) ideal and (b) non-ideal photon counting multi-bin systems. Bin edges are set to produce approximately equal number of counts in each bin for the projection data generated with the validation phantom.}\label{fig:response_functions}
			\end{figure}

			The product of detection efficiency and bin response functions for the ideal and the non-ideal multi-bin systems are presented in Fig. \ref{fig:response_functions}.

	\subsection{Energy Integrating System Image Reconstruction With Water Correction}

		We implement the emprical cupping correction of \cite{kachelriess2006empirical} in the image reconstruction of the projection data generated for the energy integrating system, which is performed as follows. Given the projection data $p$ of the calibration phantom, a set of basis images $a_0(\x),\dots,a_M(\x)$ is computed from monomials of the projection values,
		\begin{equation}
			a_m(\x) = \mathcal{R}^{-1}P_m(p)(\x), \quad P_m(p) = \left\{p_i^m\right\}_{i=1}^N.
		\end{equation}
		A set of coefficients $c_0,\dots,c_M$ are then computed such that
		\begin{equation}
			a(\x) = \sum_{m=0}^M c_m a_m(\x)
		\end{equation}
		exhibits no cupping artefacts, by minimizing the functional
		\begin{equation}
			\sum_{\x}m(\x)(a(\x) - t(\x))^2,
		\end{equation}
		where the sum is over all pixels in the image, $m(\x)$ is a weight designed not to take edges into account and $t(\x)$ is a template of the known calibration phantom Hounsfield unit values, in our case of water and air. The image reconstruction of the validation phantom is performed in the same way, with the coefficients $c_0,\dots,c_M$ found to minimize cupping for the calibration phantom.

\begin{figure}[htbp]
			\centering						
			\begin{minipage}[b]{\linewidth}
				\subfloat[\label{fig:uncorrected_water}]{\includegraphics[width=0.5\linewidth]{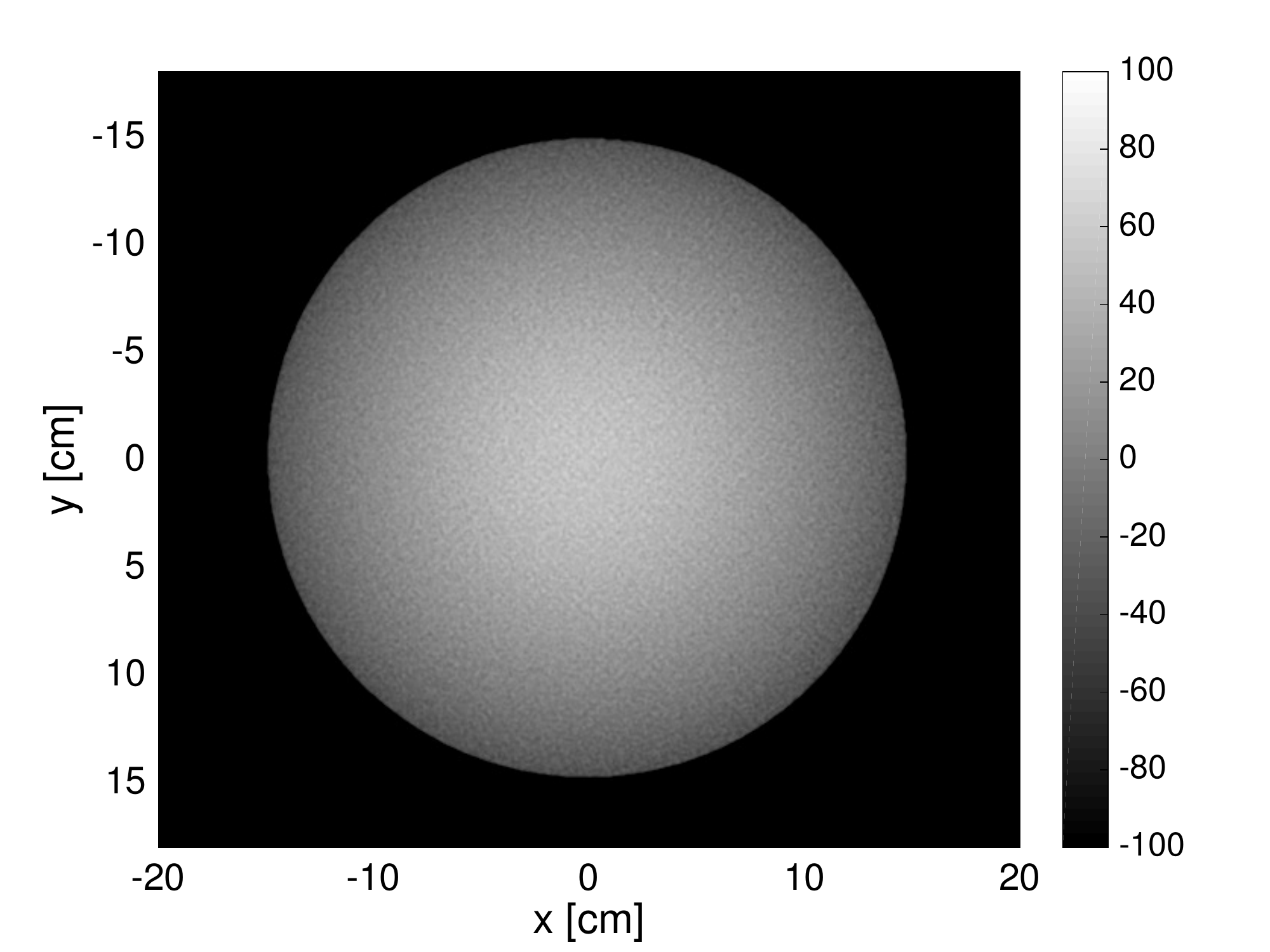}}
				\subfloat[\label{fig:corrected_water}]{\includegraphics[width=0.5\linewidth]{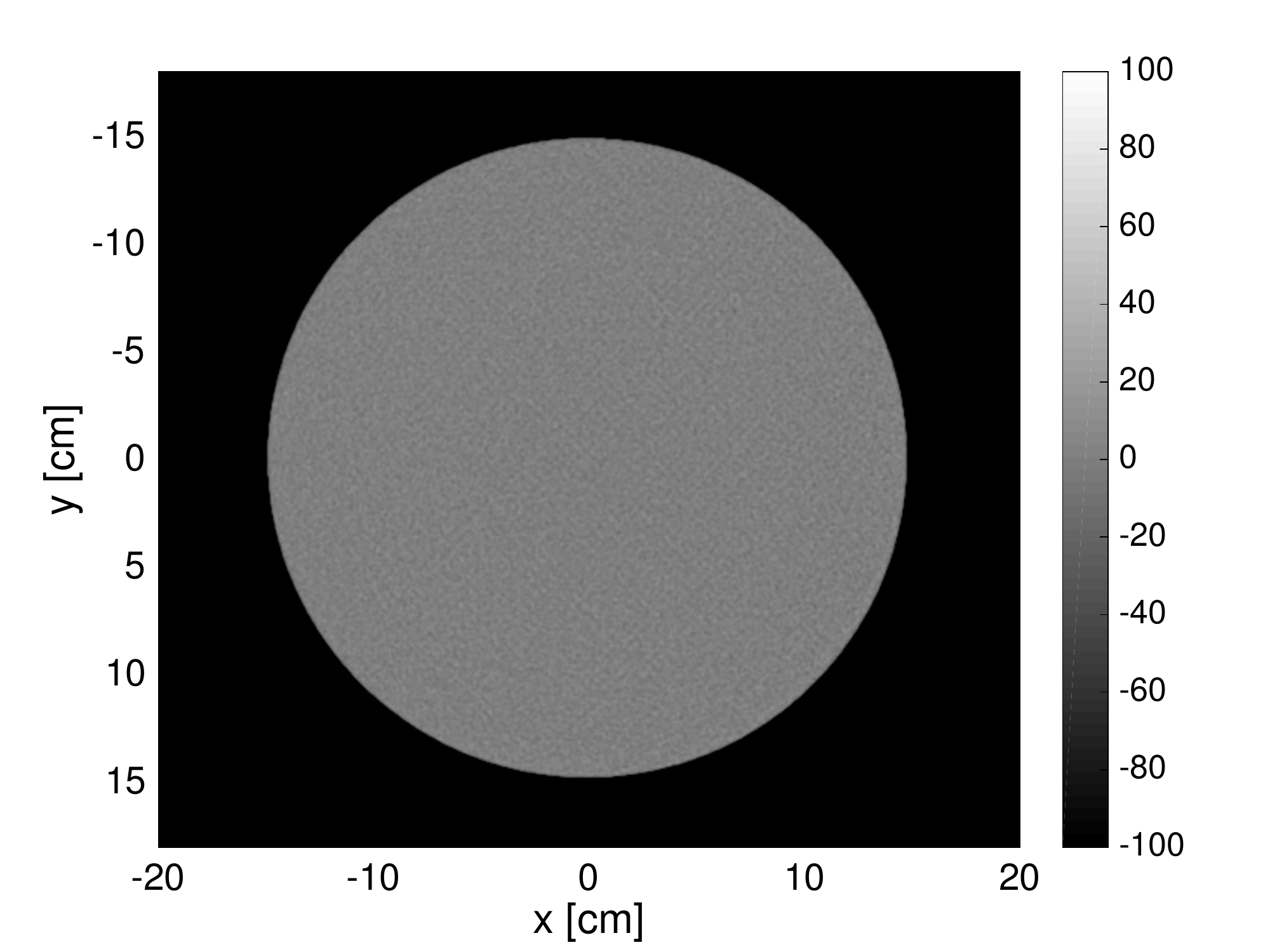}}
				\vfil
				\subfloat[\label{fig:cupping_basis}]{
					\begin{tabular}[b]{c c c}
					\includegraphics[width=0.16\linewidth]{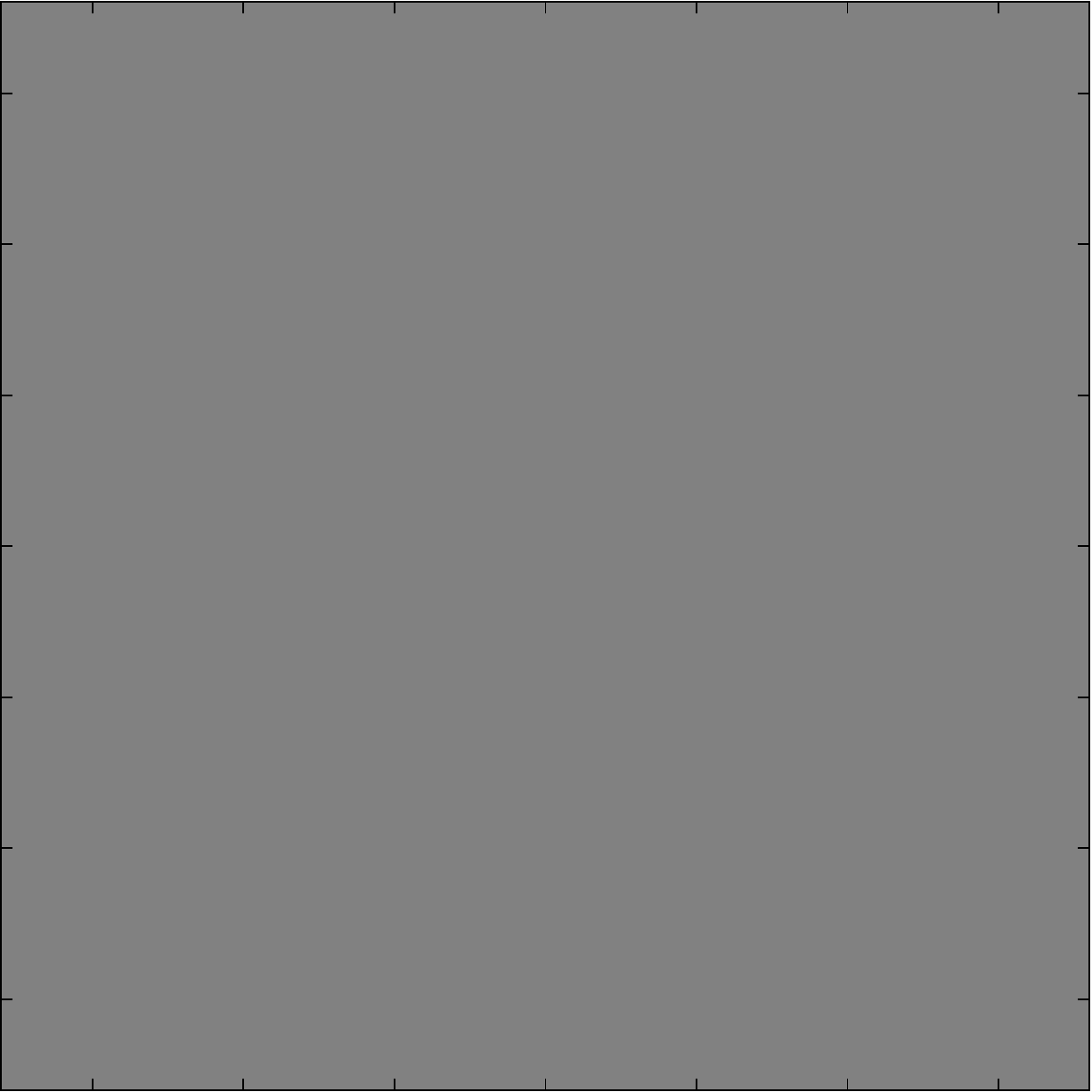}
					&\includegraphics[width=0.16\linewidth]{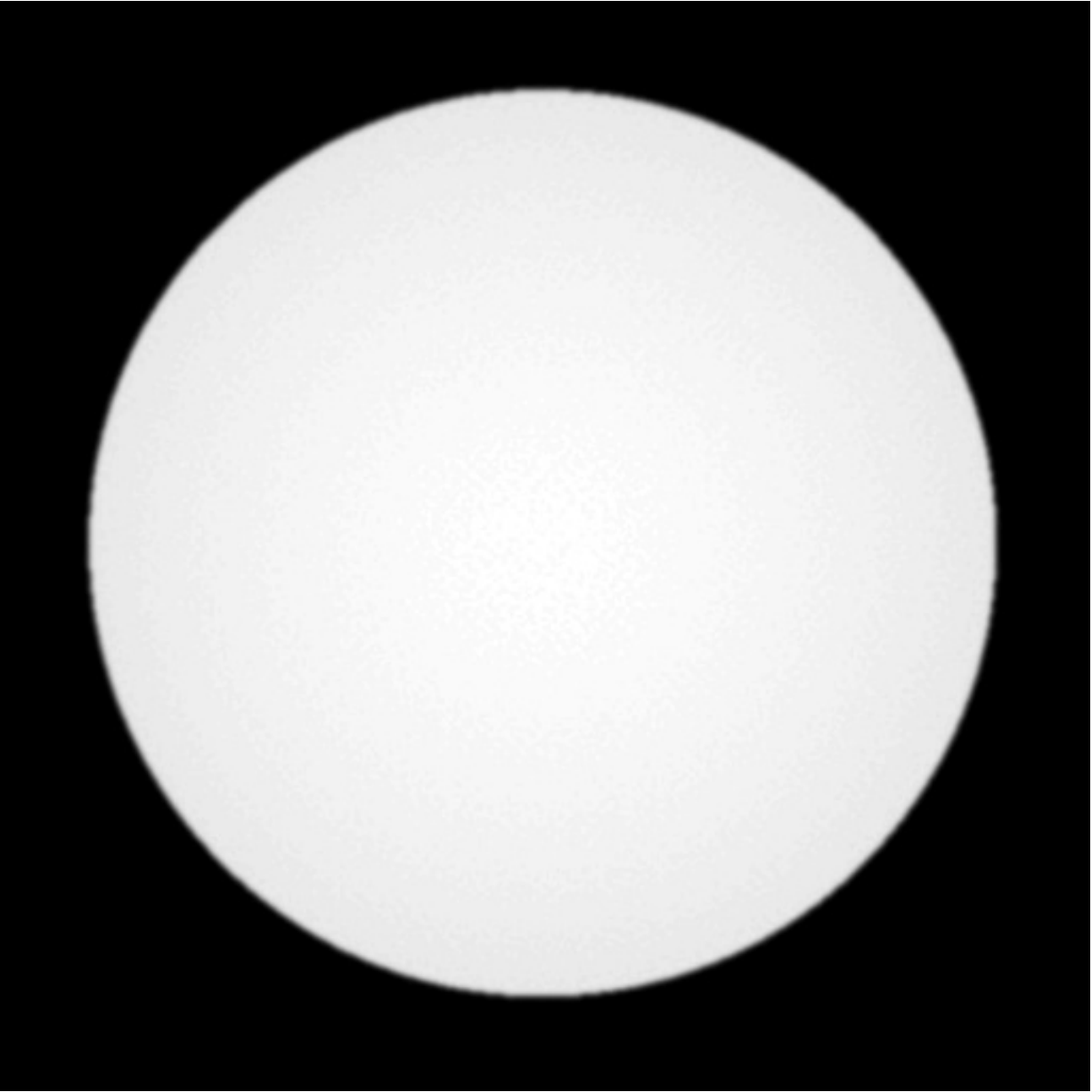}
					&\includegraphics[width=0.16\linewidth]{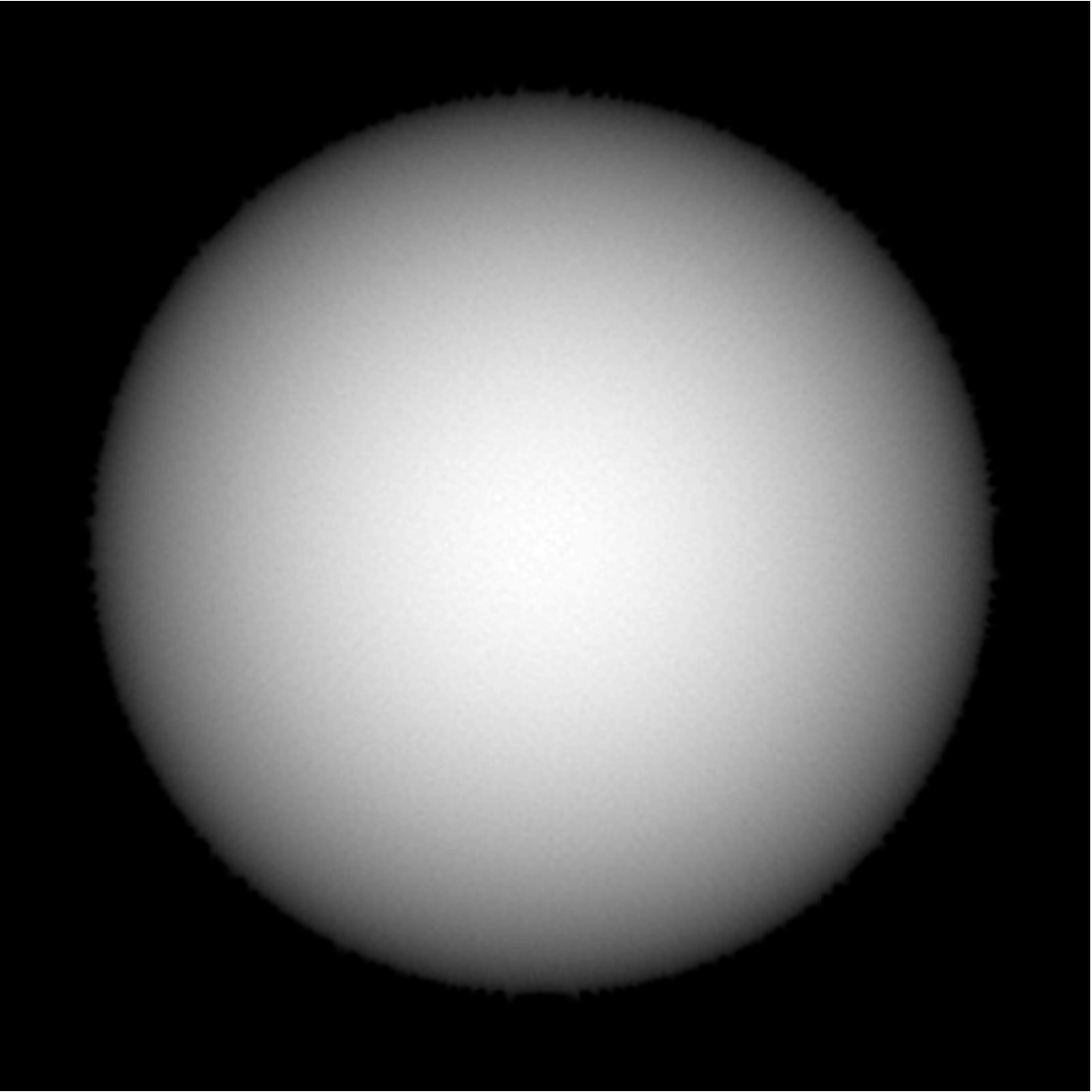}
					\\
					\includegraphics[width=0.16\linewidth]{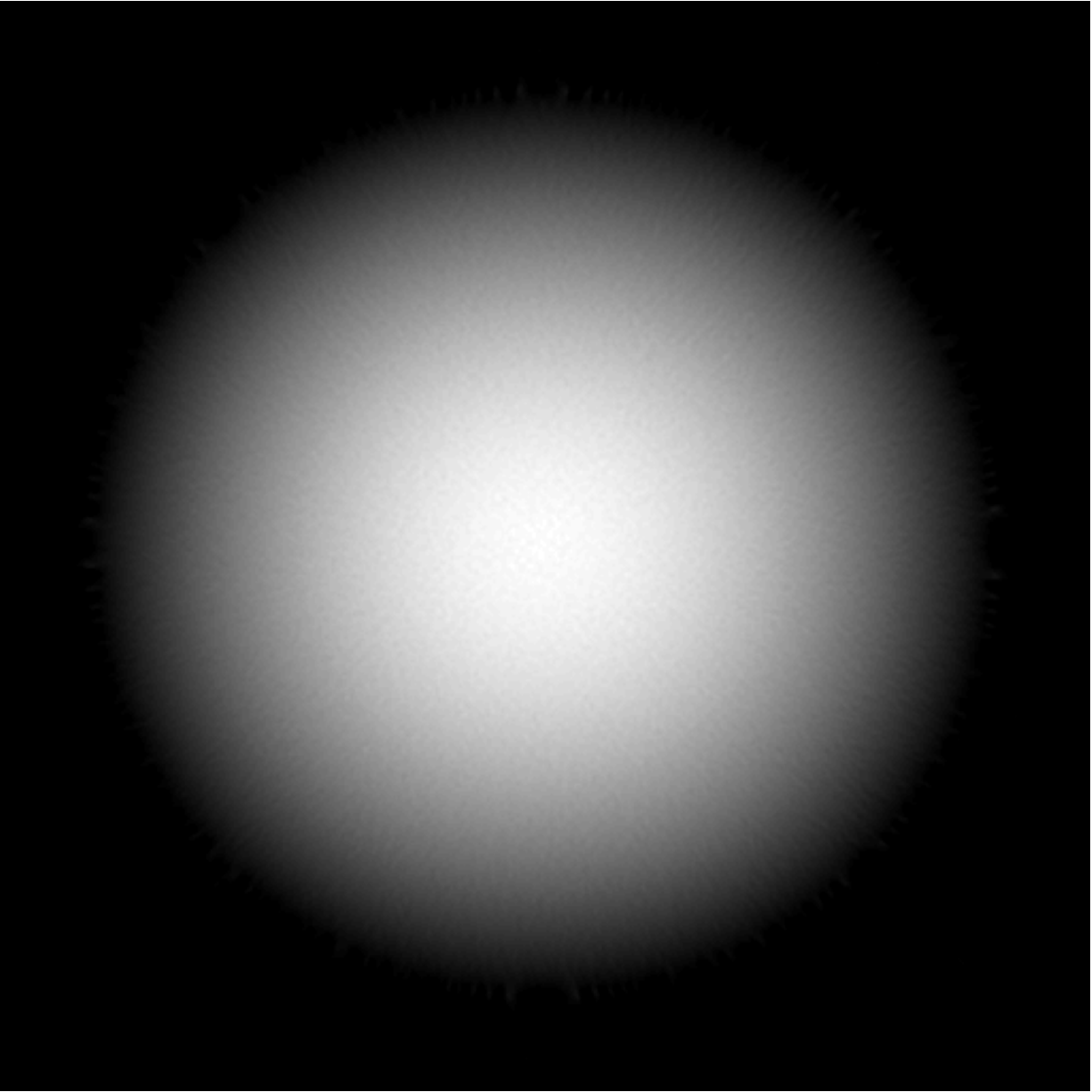}
					&\includegraphics[width=0.16\linewidth]{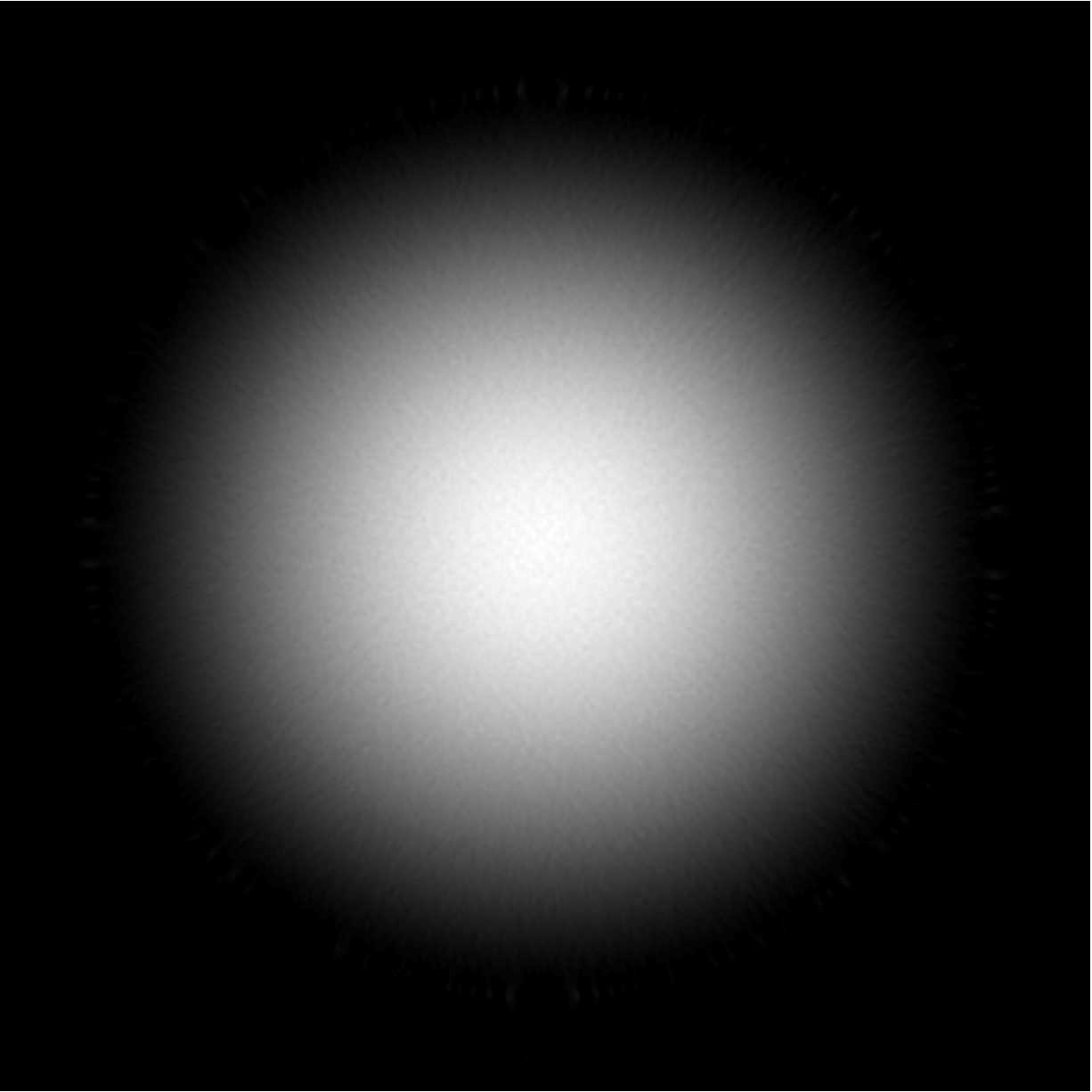}
					&\includegraphics[width=0.16\linewidth]{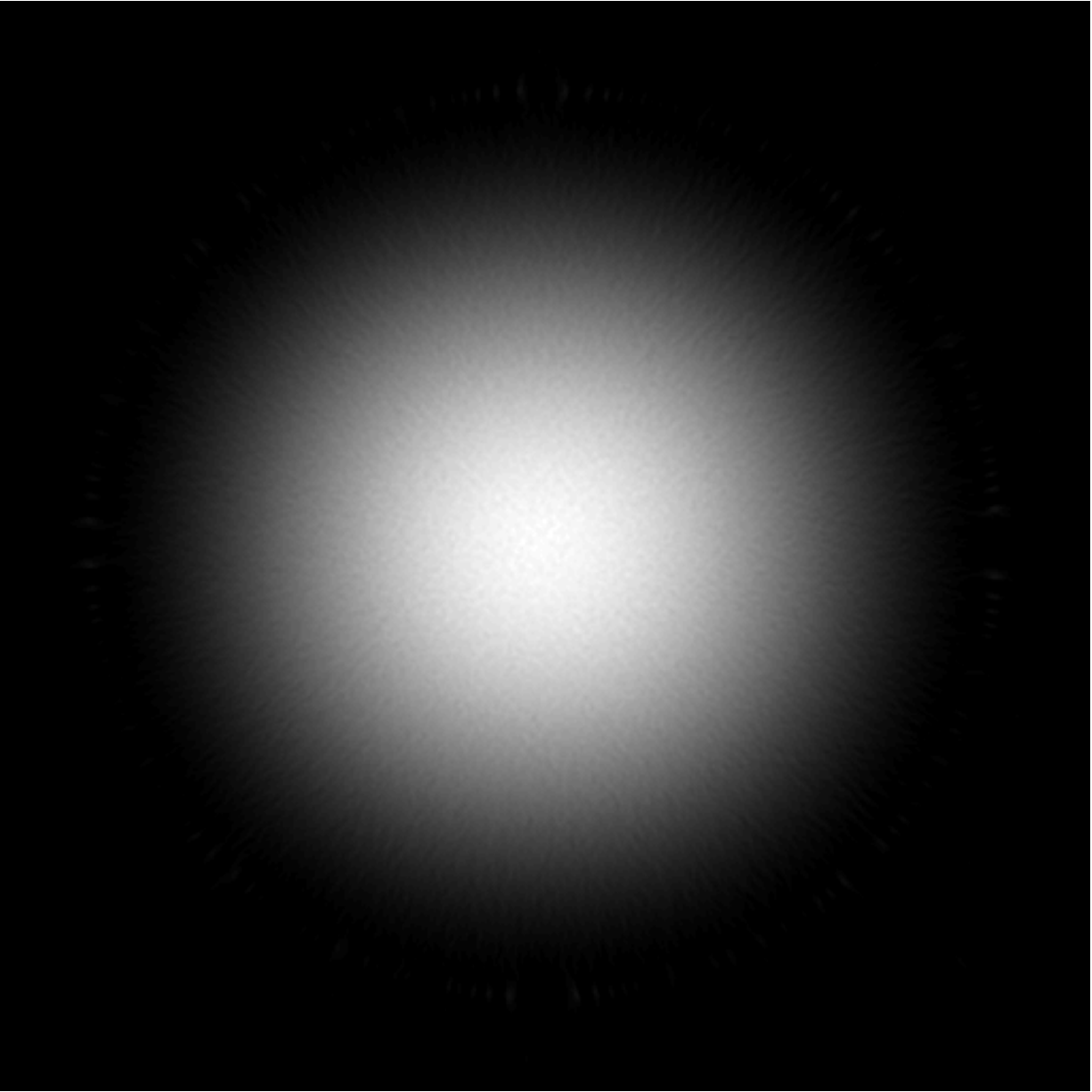}
					\end{tabular}
					}
			\end{minipage}	
			\caption{Water correction images. The uncorrected (a) and corrected (b) reconstructions are shown in a Hounsfield unit window of $[-100,100]$. Cupping correction basis images are shown in (c).}\label{fig:water_correction}					
		\end{figure}

		We use a set of six basis images, where the zeroth basis image is a constant background. The monomial basis images are shown in Fig. \ref{fig:cupping_basis}, the uncorrected reconstruction in Fig. \ref{fig:uncorrected_water} and the corrected reconstruction in Fig. \ref{fig:corrected_water}. Note that the implementation of a bowtie-filter results in capping rather than cupping, caused by more beam-hardening in non-central projection lines. The artefact is still removable by the ECC, as seen in Fig. \ref{fig:corrected_water}, a testament to the generality of the method.

	\subsection{Basis Decomposition and Creation of Synthetic Hounsfield Unit Images}

		Basis decomposition is performed by solving the inequality constrained optimization problem (\ref{eq:ML}) for the generated projection data sets using a \textsc{Matlab} implementation of the barrier method with Newton's method as the inner solver, as described in \cite{boyd2009convex}. The LACs of water and bone are used as basis functions. Basis images $\bhata(\x)$ are created according to \eqref{eq:image_estimates} using \textsc{Matlab}'s \texttt{ifanbeam} implementation of filtered back projection with a cosine filter.

		Let $\bw$ denote a normalized discretization of the weighting function \eqref{eq:ideal_weight}, used to compute the ideal Hounsfield units presented in Tab. \ref{tab:materials}, i.e. The coefficient vector $\bhatf_w$ used to create ideal effective LACs is computed according to \eqref{eq:effective_decomposition},
		\begin{equation}
			\bhatf_w = \bw^T\!\bef \in \R^{1\times L},
		\end{equation}
		where $\bef$ is the matrix of discretized basis functions used in the basis decomposition. Synthetic Hounsfield unit images are then computed according to \eqref{eq:SHU}, with estimated basis coefficients of air and water determined from the reconstruction of the calibration phantom.

	\subsection{Figure of Merit}

		The CT number accuracy in the reconstructed images are computed with respect to the ideal Hounsfield units presented in Tab. \ref{tab:materials}. We use the sample bias, sample standard deviation and root-mean-squared error (RMSE) within a region of interest for each material in the validation phantom as figures of merit. The regions of interest are shown in Fig. \ref{fig:roi} and do not contain the boundary region between materials.

\section{Results}

\begin{figure}[htbp]
		\centering						
		\begin{minipage}[b]{\linewidth}
			\subfloat[]{\includegraphics[width=0.5\linewidth]{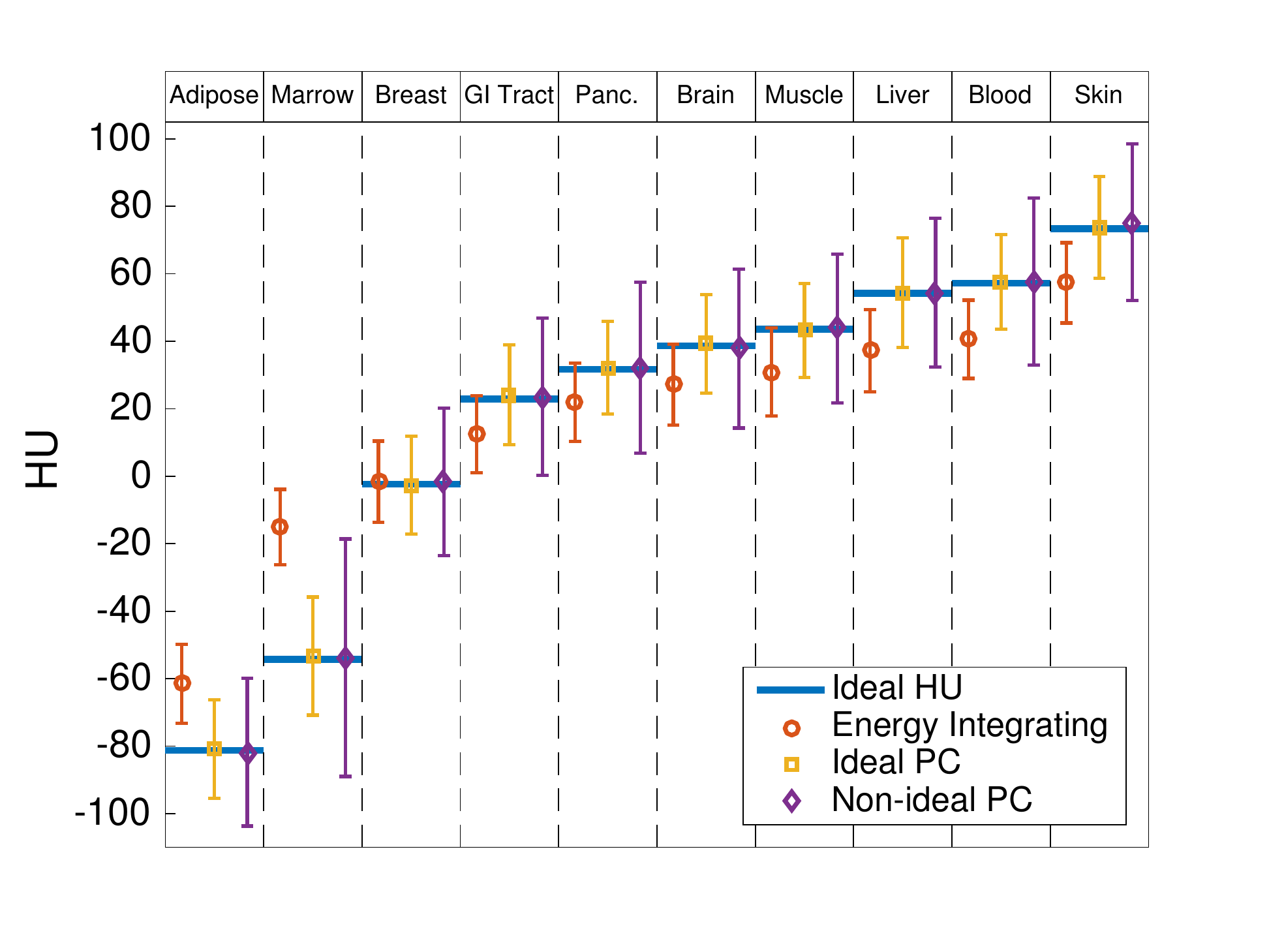}}
			\subfloat[]{\includegraphics[width=0.5\linewidth]{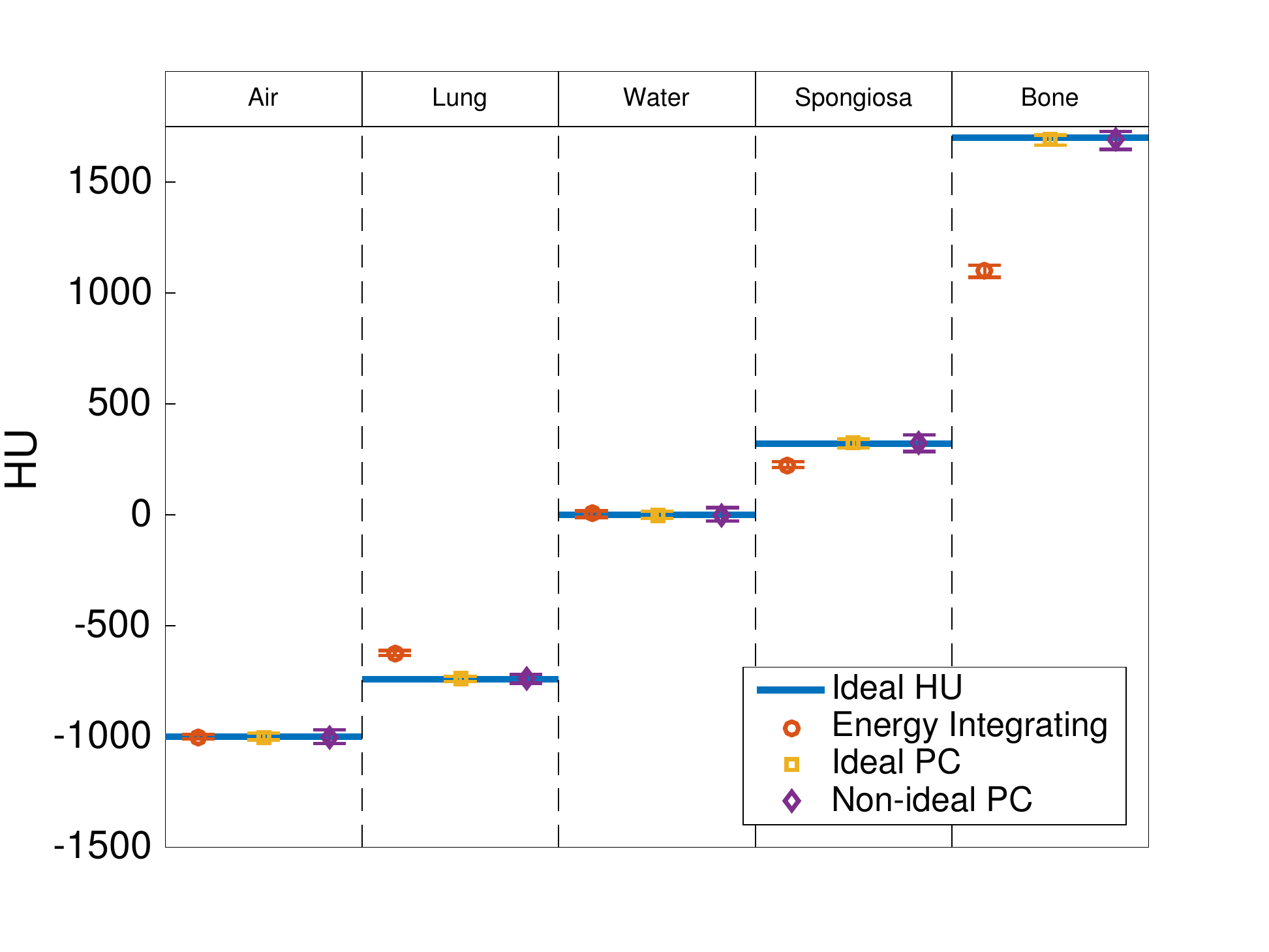}}			
		\end{minipage}
		\caption{Sample mean and error bars of one standard deviation of reconstructed CT numbers for each material in the validation phantom, computed in regions of interest shown in Fig. \ref{fig:roi}, for each simulated system. (a) Sample mean and error for tissues with HU values between -100 and 100. (b) Sample mean and error for tissues with HU values smaller than -100 or greater than 100, as well as air and water.}\label{fig:error_plot}
	\end{figure}

\begin{figure}[htbp]
		\centering						
		\begin{minipage}[b]{\linewidth}
			\subfloat[]{\includegraphics[width=0.44\linewidth]{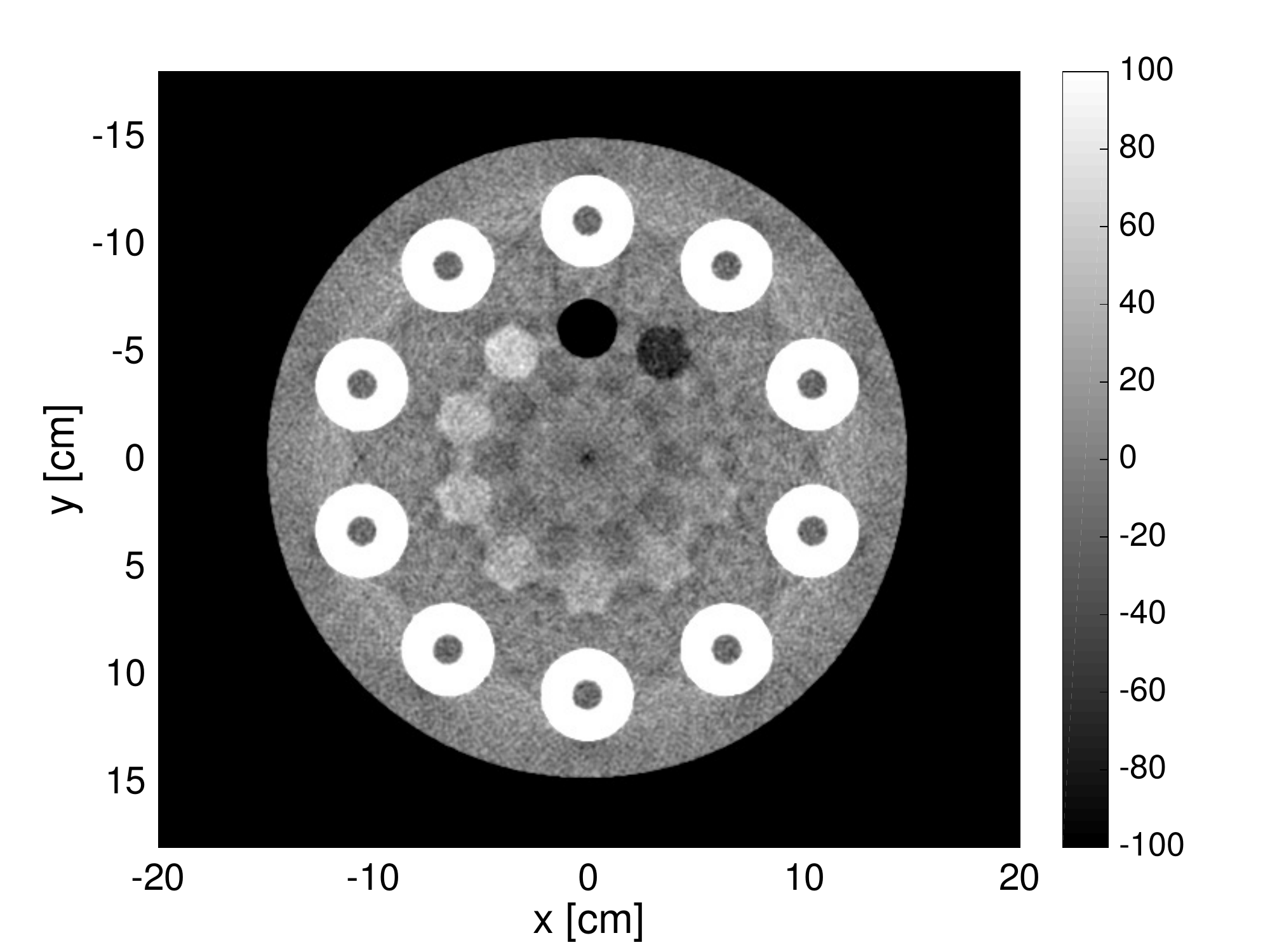}}
			\hfil
			\subfloat[]{\includegraphics[width=0.44\linewidth]{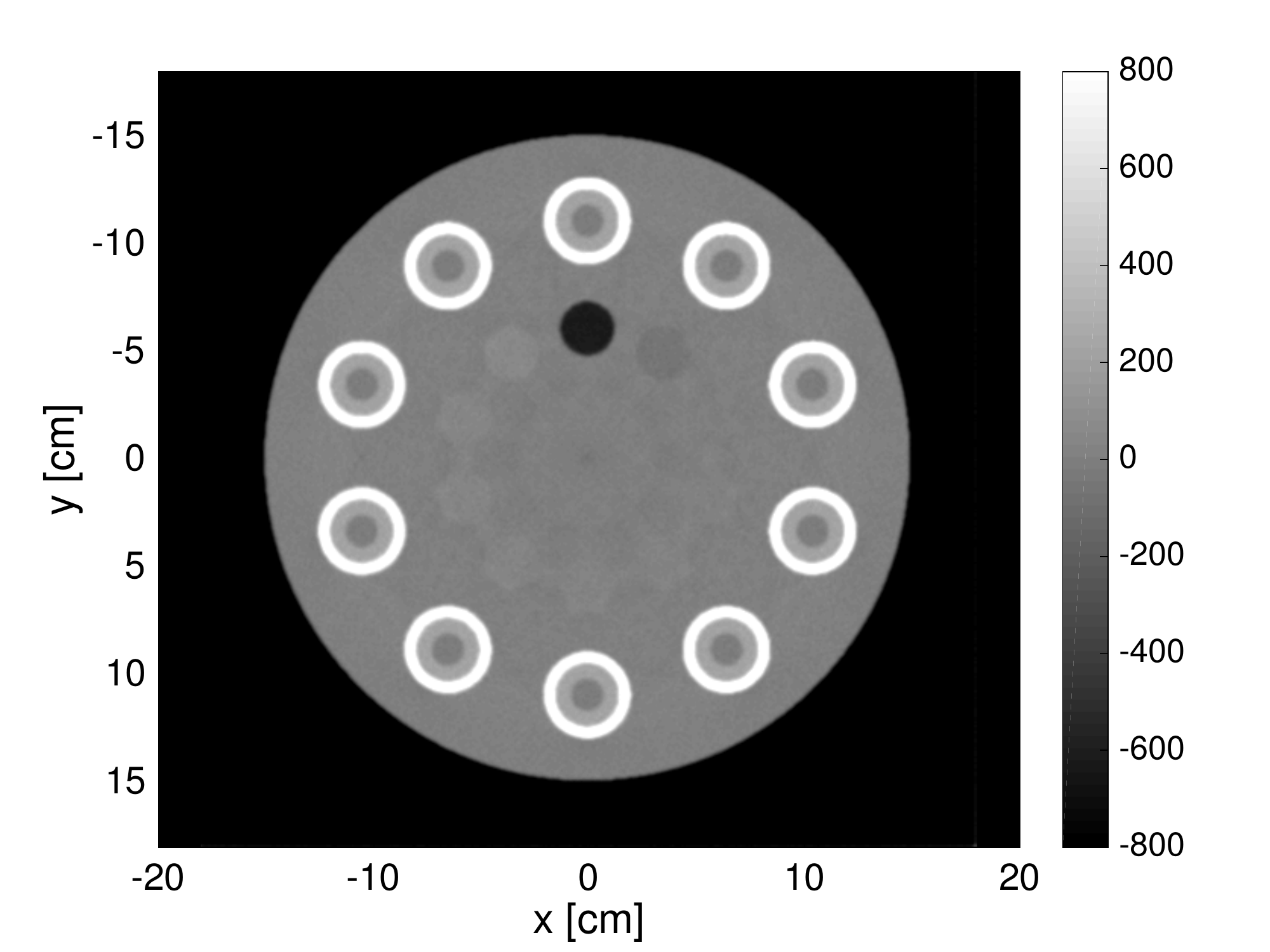}}				
			\vfil
			\subfloat[]{\includegraphics[width=0.44\linewidth]{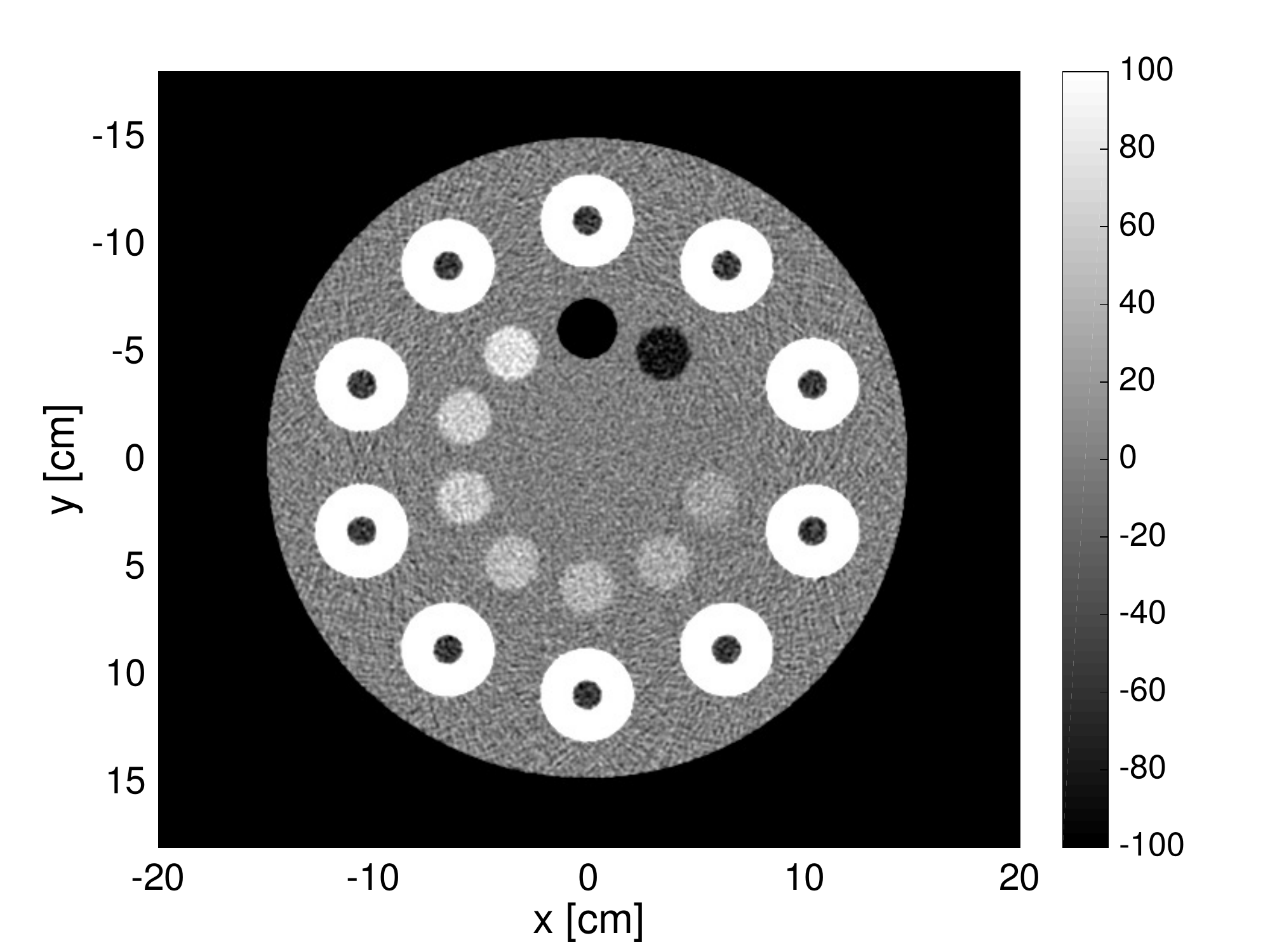}}
			\hfil
			\subfloat[]{\includegraphics[width=0.44\linewidth]{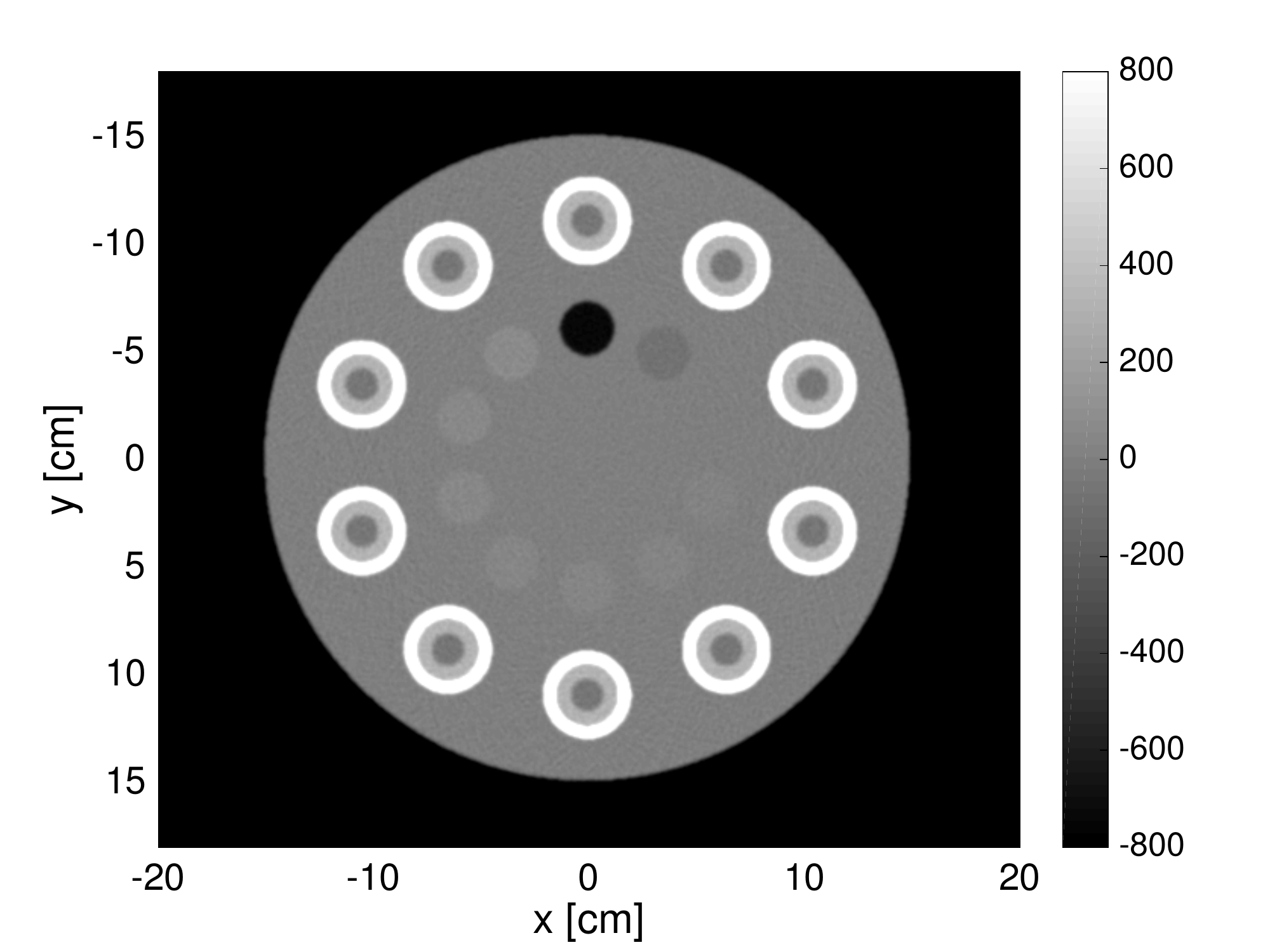}}	
			\vfil
			\subfloat[]{\includegraphics[width=0.44\linewidth]{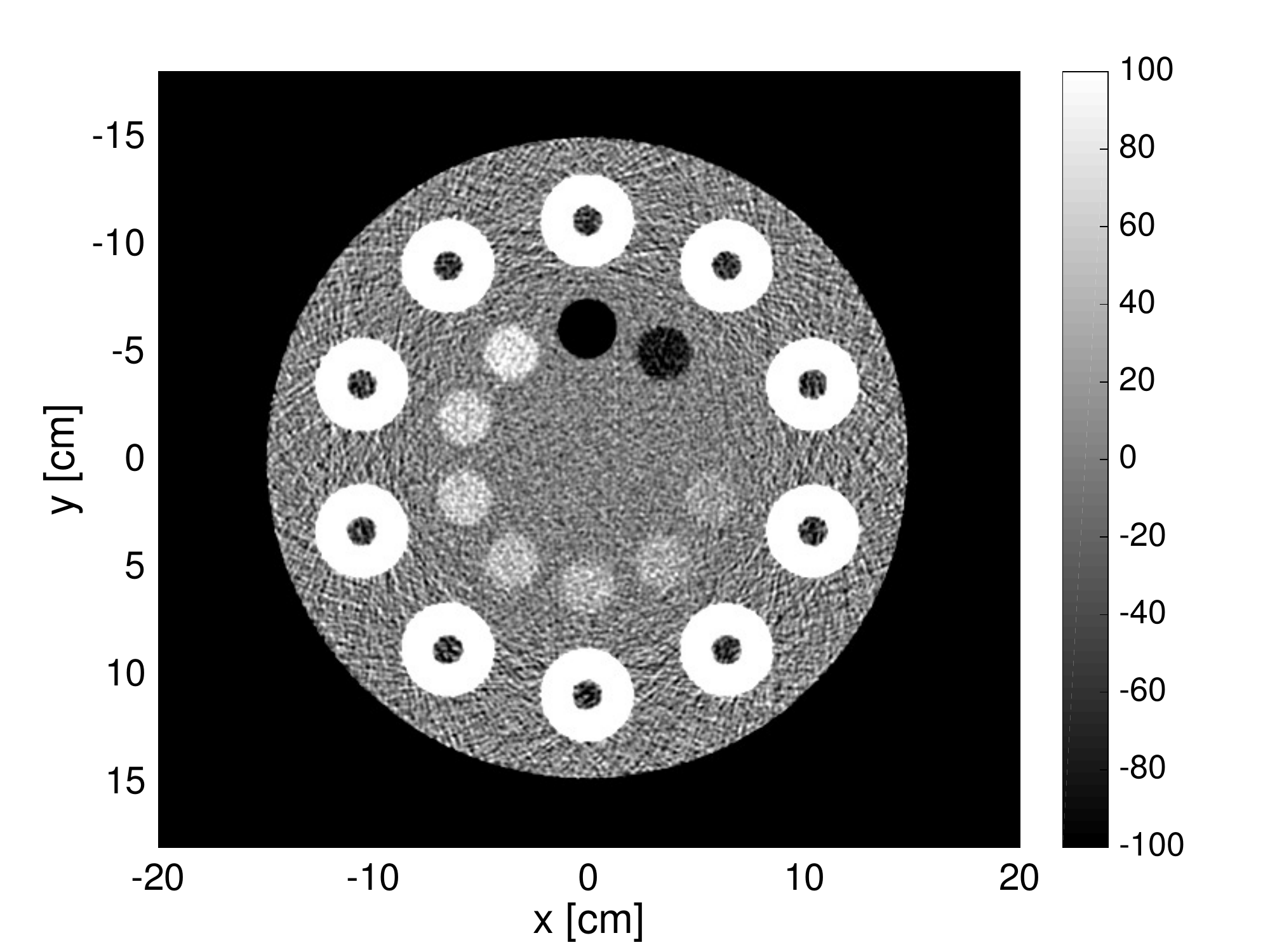}}
			\hfil
			\subfloat[]{\includegraphics[width=0.44\linewidth]{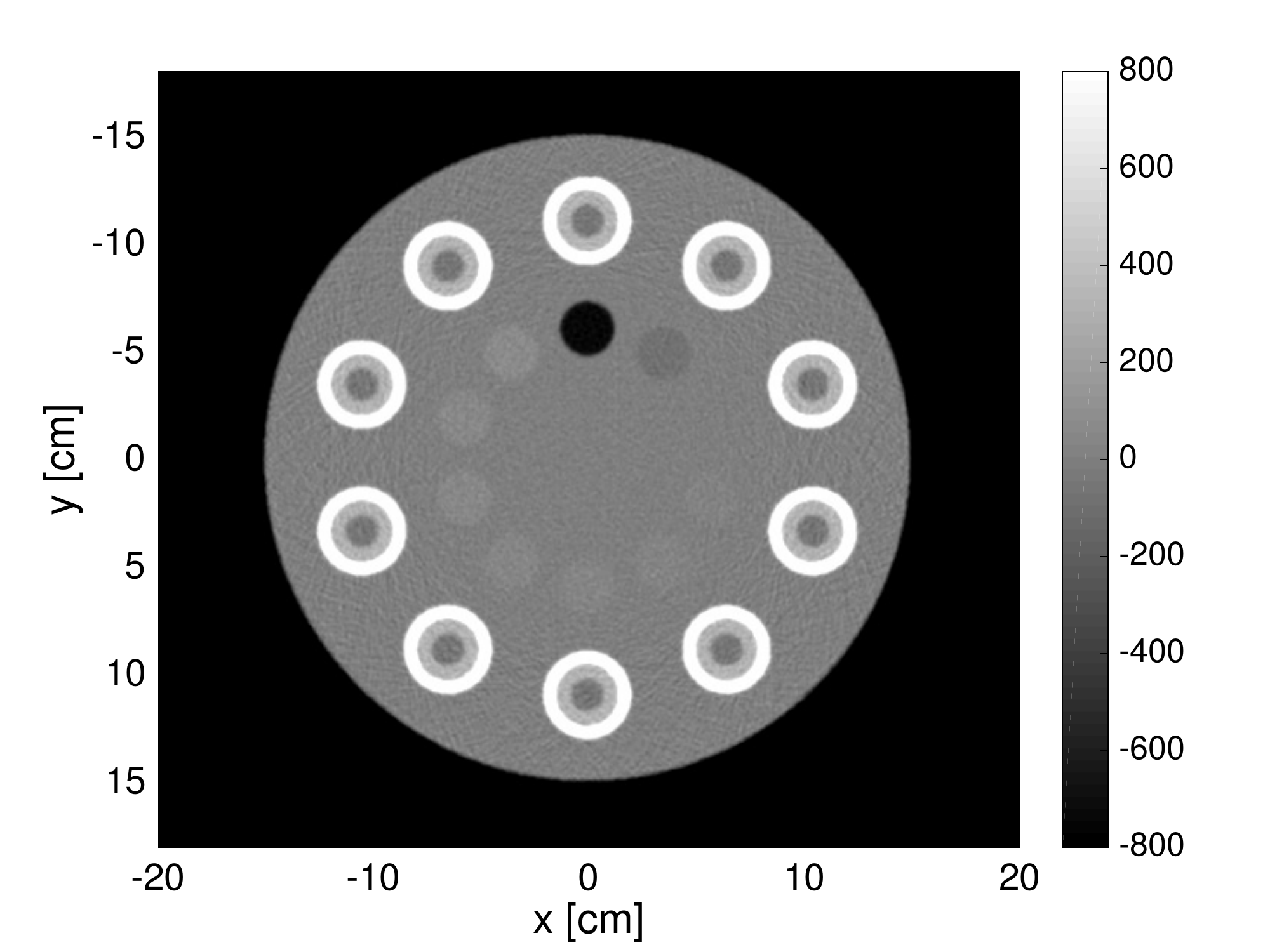}}		
		\end{minipage}
		\caption{Reconstructions of the validation phantom for the simulated systems; ideal energy integrating (top row), ideal multi-bin (middle row) and non-ideal multi-bin (bottom row). The images are displayed in Hounsfield unit windows of [-100,100] (left column) and [-800,800] (right column).}\label{fig:reconstructions}
	\end{figure}

	Reconstructions from the three simulated systems are shown in Fig. \ref{fig:reconstructions}. The sample bias and RMSE of the reconstructed CT numbers for each material in the validation phantom are presented in Tab. \ref{tab:results}. An error bar plot of the reconstructed CT numbers is shown in Fig. \ref{fig:error_plot}.

	\begin{table}[htbp]
		\centering
		\begin{tabular}{r | r | r | r | r | r | r}
		& \multicolumn{2}{c|}{Conventional} & \multicolumn{2}{c|}{Ideal MB} & \multicolumn{2}{c}{Non-ideal MB} \\
		\hline
		Material 		& \multicolumn{1}{c|}{Bias} 		& \multicolumn{1}{c|}{RMSE}	 	& \multicolumn{1}{c|}{Bias} 	& \multicolumn{1}{c|}{RMSE} 	& \multicolumn{1}{c|}{Bias} 	& \multicolumn{1}{c}{RMSE}	\\
		\hline
		Air & -0.1 & 9.9 & -0.2 & 15.3 & -0.8 & 31.4 \\
		Water & 2.8 & 15.0 & 0.6 & 16.4 & 1.7 & 30.7 \\
		\hline
		Yel. Marrow & 39.2 & 40.8 & 1.0 & 17.5 & 0.5 & 35.2 \\
		Spongiosa & -94.6 & 95.3 & 1.0 & 19.5 & 1.8 & 37.2 \\
		Bone & -602.3 & 603.0 & -10.9 & 24.6 & -12.1 & 42.6 \\
		\hline
		Inflated Lung & 118.0 & 118.5 & 1.3 & 12.5 & 2.1 & 19.9 \\
		Adipose & 19.7 & 22.9 & 0.5 & 14.6 & -0.5 & 21.9 \\
		Breast & 0.7 & 12.1 & -0.3 & 14.5 & 0.6 & 21.9 \\
		GI Tract & -10.5 & 15.5 & 1.2 & 14.8 & 0.7 & 23.3 \\
		Pancreas & -9.8 & 15.2 & 0.5 & 13.7 & 0.5 & 25.3 \\
		Brain & -11.4 & 16.6 & 0.6 & 14.6 & -0.8 & 23.5 \\
		Muscle & -12.7 & 18.2 & -0.4 & 13.9 & 0.2 & 22.1 \\
		Liver & -17.0 & 20.9 & 0.2 & 16.2 & 0.2 & 21.9 \\
		Blood & -16.6 & 20.3 & 0.4 & 14.0 & 0.5 & 24.8 \\
		Skin & -16.1 & 20.1 & 0.3 & 15.1 & 1.9 & 23.2
		\end{tabular}\caption{Sample bias and RMSE of the reconstructed CT numbers for the conventional energy integrating system, the ideal multi-bin system (Ideal MB) and the non-ideal multi-bin system (Non-ideal MB).}\label{tab:results}
	\end{table}

	\begin{figure}[htbp]
		\centering
		\begin{minipage}[b]{\linewidth}
			\centering
			\includegraphics[width=0.45\linewidth]{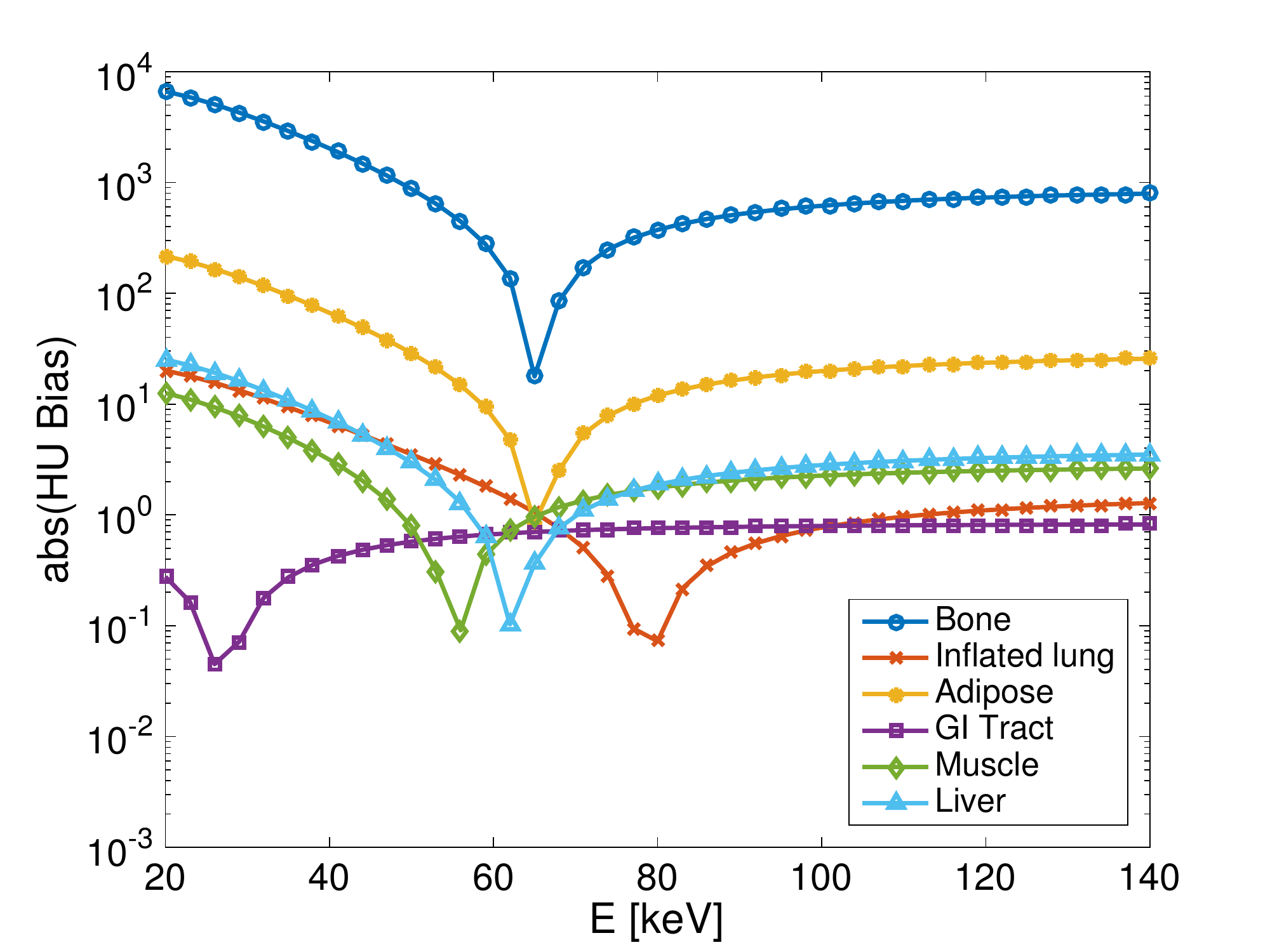}
		\end{minipage}
		\caption{Mean error of reconstructed HU values for each material in the validation phantom in a synthetic monoenergetic reconstruction from the ideal multi-bin system projection data.}\label{fig:mono_error}
	\end{figure}

\section{Conclusion}


	From the results presented in Tab. \ref{tab:results} and Fig. \ref{fig:error_plot} we conclude that the method is able to reproduce unbiased estimates of the ideal Hounsfield units defined in Section \ref{sec:ideal_units} for both of the simulated multi-bin systems, consistently for the range of tissues used in the phantom. A visual comparison of the reconstructions in Fig. \ref{fig:reconstructions} shows the synthetic Hounsfield unit images to be slightly noisier than the conventional image, but more importantly, free of artefacts.

	The method is easily adapted to mimic systems with different parameters, e.g. x-ray tube filtration or acceleration voltage, than the ones used to simulate projection data, by choosing an appropriate reference spectrum to compute the mapping between basis images and synthetic Hounsfield units. It is, in our opinion, a more elegant and physics based approach than e.g. image blending techniques \cite{holmes2008evaluation}.


	Although beam hardening mitigators such as an ideal bowtie-filter and a cupping correction algorithm were implemented, the conventional system suffers from beam hardening caused by the high density bone inserts. The resulting bias is in the range between -17 HU (for liver) to +39 HU (for yellow marrow) for water-like tissues. Bone and spongiosa, because of their intrinsic beam hardening show much greater bias. The goal of the method, to produce CT numbers appearing like experienced radiologists would expect them to, might therefore not be achieved for these tissues. We do not believe this is a significant shortcoming, since CT numbers for dense tissues are also highly sensitive to system parameters and probably differ greatly on different systems.


	Basis decomposition allows for complete removal of beam hardening artefacts, which works very well in simulation. In practice, accurate forward model specification is a challenge and will cause bias in the basis coefficient estimates if not properly done. Relevant future work is therefore experimental validation of the method.

\section{Disclosure}

Fredrik Grönberg discloses past financial interests in Prismatic Sensors AB and is currently employed by GE Healthcare. Hans Bornefalk discloses past financial interests in Prismatic Sensors AB.

\section*{References}

\bibliographystyle{spiejour}


\begin{thebibliography}{10}

	\bibitem{shefer2013state}
	E.~Shefer, A.~Altman, R.~Behling, {\em et~al.}, ``State of the art of ct
	  detectors and sources: a literature review,'' {\em Current Radiology Reports}
	  {\bf 1}(1), 76--91  (2013).

	\bibitem{shikhaliev2005photon}
	P.~M. Shikhaliev, T.~Xu, and S.~Molloi, ``Photon counting computed tomography:
	  concept and initial results,'' {\em Medical physics} {\bf 32}(2), 427--436
	  (2005).

	\bibitem{schlomka2008experimental}
	J.~Schlomka, E.~Roessl, R.~Dorscheid, {\em et~al.}, ``Experimental feasibility
	  of multi-energy photon-counting k-edge imaging in pre-clinical computed
	  tomography,'' {\em Physics in medicine and biology} {\bf 53}(15), 4031
	  (2008).

	\bibitem{iwanczyk2009photon}
	J.~S. Iwanczyk, E.~Nyg{\aa}rd, O.~Meirav, {\em et~al.}, ``Photon counting
	  energy dispersive detector arrays for x-ray imaging,'' {\em Nuclear Science,
	  IEEE Transactions on} {\bf 56}(3), 535--542  (2009).

	\bibitem{bornefalk2010photon}
	H.~Bornefalk and M.~Danielsson, ``Photon-counting spectral computed tomography
	  using silicon strip detectors: a feasibility study,'' {\em Physics in
	  medicine and biology} {\bf 55}(7), 1999  (2010).

	\bibitem{kappler2012first}
	S.~Kappler, T.~Hannemann, E.~Kraft, {\em et~al.}, ``First results from a hybrid
	  prototype ct scanner for exploring benefits of quantum-counting in clinical
	  ct,'' in {\em SPIE Medical Imaging},  83130X--83130X, International Society
	  for Optics and Photonics  (2012).

	\bibitem{alvarez1976energy}
	R.~E. Alvarez and A.~Macovski, ``Energy-selective reconstructions in x-ray
	  computerised tomography,'' {\em Physics in medicine and biology} {\bf 21}(5),
	  733  (1976).

	\bibitem{alvarez1979comparison}
	R.~Alvarez and E.~Seppi, ``A comparison of noise and dose in conventional and
	  energy selective computed tomography,'' {\em Nuclear Science, IEEE
	  Transactions on} {\bf 26}(2), 2853--2856  (1979).

	\bibitem{lehmann1981generalized}
	L.~Lehmann, R.~Alvarez, A.~Macovski, {\em et~al.}, ``Generalized image
	  combinations in dual kvp digital radiography,'' {\em Medical physics} {\bf
	  8}(5), 659--667  (1981).

	\bibitem{roessl2007k}
	E.~Roessl and R.~Proksa, ``K-edge imaging in x-ray computed tomography using
	  multi-bin photon counting detectors,'' {\em Physics in medicine and biology}
	  {\bf 52}(15), 4679  (2007).

	\bibitem{flohr2006first}
	T.~G. Flohr, C.~H. McCollough, H.~Bruder, {\em et~al.}, ``First performance
	  evaluation of a dual-source ct (dsct) system,'' {\em European radiology} {\bf
	  16}(2), 256--268  (2006).

	\bibitem{kalender1986evaluation}
	W.~A. Kalender, W.~Perman, J.~Vetter, {\em et~al.}, ``Evaluation of a prototype
	  dual-energy computed tomographic apparatus. i. phantom studies,'' {\em
	  Medical physics} {\bf 13}(3), 334--339  (1986).

	\bibitem{goodsitt2011accuracies}
	M.~M. Goodsitt, E.~G. Christodoulou, and S.~C. Larson, ``Accuracies of the
	  synthesized monochromatic ct numbers and effective atomic numbers obtained
	  with a rapid kvp switching dual energy ct scanner,'' {\em Medical physics}
	  {\bf 38}(4), 2222--2232  (2011).

	\bibitem{taguchi2007image}
	K.~Taguchi, M.~Zhang, E.~C. Frey, {\em et~al.}, ``Image-domain material
	  decomposition using photon-counting ct,'' in {\em Medical Imaging},
	  651008--651008, International Society for Optics and Photonics  (2007).

	\bibitem{holmes2008evaluation}
	D.~R. Holmes, J.~G. Fletcher, A.~Apel, {\em et~al.}, ``Evaluation of non-linear
	  blending in dual-energy computed tomography,'' {\em European journal of
	  radiology} {\bf 68}(3), 409--413  (2008).

	\bibitem{kim2010image}
	K.~S. Kim, J.~M. Lee, S.~H. Kim, {\em et~al.}, ``Image fusion in dual energy
	  computed tomography for detection of hypervascular liver hepatocellular
	  carcinoma: phantom and preliminary studies,'' {\em Investigative radiology}
	  {\bf 45}(3), 149--157  (2010).

	\bibitem{bornefalk2012synthetic}
	H.~Bornefalk, ``Synthetic hounsfield units from spectral ct data,'' {\em
	  Physics in medicine and biology} {\bf 57}(7), N83  (2012).

	\bibitem{kachelriess2006empirical}
	M.~Kachelrie{\ss}, K.~Sourbelle, and W.~A. Kalender, ``Empirical cupping
	  correction: a first-order raw data precorrection for cone-beam computed
	  tomography,'' {\em Medical physics} {\bf 33}(5), 1269--1274  (2006).

	\bibitem{roessl2009cramer}
	E.~Roessl and C.~Herrmann, ``Cram{\'e}r--rao lower bound of basis image noise
	  in multiple-energy x-ray imaging,'' {\em Physics in medicine and biology}
	  {\bf 54}(5), 1307  (2009).

	\bibitem{berger2013xcom}
	M.~Berger, J.~Hubbell, S.~Seltzer, {\em et~al.}, ``Xcom: Photon cross sections
	  database,'' {\em NIST Standard reference database} {\bf 8}  (2013).

	\bibitem{white1989tissue}
	D.~White, J.~Booz, R.~Griffith, {\em et~al.}, ``Tissue substitutes in radiation
	  dosimetry and measurement,'' {\em ICRU Report} {\bf 44}  (1989).

	\bibitem{cranley1997catalogue}
	K.~Cranley, B.~Gilmore, G.~Fogarty, {\em et~al.}, ``Catalogue of diagnostic
	  x-ray spectra and other data,'' {\em IPEM report} {\bf 78}  (1997).

	\bibitem{klein1929streuung}
	O.~Klein and Y.~Nishina, ``{\"U}ber die streuung von strahlung durch freie
	  elektronen nach der neuen relativistischen quantendynamik von dirac,'' {\em
	  Zeitschrift f{\"u}r Physik} {\bf 52}(11-12), 853--868  (1929).

	\bibitem{xu2013evaluation}
	C.~Xu, M.~Persson, H.~Chen, {\em et~al.}, ``Evaluation of a second-generation
	  ultra-fast energy-resolved asic for photon-counting spectral ct,'' {\em
	  Nuclear Science, IEEE Transactions on} {\bf 60}(1), 437--445  (2013).

	\bibitem{boyd2009convex}
	S.~Boyd and L.~Vandenberghe, {\em Convex optimization}, Cambridge university
	  press  (2009).

\end{thebibliography}

\end{document}